\documentclass[onecolumn,showpacs,showkeys,preprintnumbers,amsmath,amssymb,floatfix]{revtex4}

\usepackage{graphicx}
\usepackage{epsfig}
\usepackage{color}
\usepackage{dsfont}
\usepackage{dcolumn}
\usepackage{bm}

\usepackage{hyperref}

\begin{document}
\title{Low and high spin mesons from $N_f=2$ Clover-Wilson lattices}
\author{Tommy Burch$^1$, 
Christian Hagen$^1$,
Martin Hetzenegger$^1$, and 
Andreas Sch\"afer$^1$}
\vskip1mm
\affiliation{$^1$Institut f\"ur Theoretische Physik, Universit\"at Regensburg, 
D-93040 Regensburg, Germany}
\begin{abstract}
We present results for excited meson spectra from $N_f=2$ clover-Wilson 
configurations provided by the CP-PACS Collaboration. In our study we investigate 
both low and high spin mesons. For spin-$0$ and spin-$1$ mesons, we are 
especially interested in the excited states. To access these states we 
construct several different interpolators from quark sources of different 
spatial smearings and calculate a matrix of correlators. For this matrix we then
solve a generalized eigenvalue problem. For spin-$2$ and spin-$3$, we 
extract only the lowest lying states. 
\end{abstract}
\maketitle

\section{Introduction}\label{SectIntroduction}

The calculation of hadron masses is one of the central subjects in lattice
QCD since it gives us the opportunity to study such nonperturbative quantities 
from first principles. The results of such calculations (with their proper 
extrapolations) can then be compared directly to experiment. This allows us to 
clarify the internal structure of experimentally known resonances and also enables us
to predict masses and properties of states which have not yet been found.
Since the precise nature of many hadron resonances is unknown, lattice QCD
calculations provide an indispensable contribution to their understanding.

However, this is not the only reason why hadron masses are the subject of
very intensive studies in lattice QCD. A second, more technical reason is
that we want to know to what extent our calculations are affected by
systematic errors, which are usually connected to limited computer
resources. The calculation of hadron masses gives us the possibility to
study these systematics of our formulation by allowing us to compare our
results directly with precise experimental measurements.

While it is well understood how to extract the mass of the ground state in 
a given channel, a clean extraction of the masses of excited states in a 
lattice QCD calculation is still a challenge. One of the main difficulties 
is the fact that excited states only appear as subleading exponentials in 
Euclidean two-point correlation functions. To extract them, a variety of 
approaches has been tried. They reach from brute-force multiexponential 
fits \cite{Aubin:2004wf} to more sophisticated techniques using Bayesian priors methods 
\cite{Asakawa:2000tr,Lepage:2001ym,Chen:2004gp} and ``NMR-inspired blackbox`` 
methods \cite{Fleming:2004hs}. Even evolutionary algorithms have been 
considered \cite{vonHippel:2007dz}. A number of these methods have been 
studied and compared in \cite{Lin:2007iq}. However, probably the most 
powerful technique is the variational approach 
\cite{Michael:1985ne,Luscher:1990ck,Burch:2005wd,Blossier:2008tx,Blossier:2009kd}, 
which is also the method we use for our studies. In this approach one studies not 
only a single correlator but a whole matrix of correlation functions.

To access the crucial information contained in this matrix, a rich enough 
basis of interpolating operators (i.e., products of creation and annihilation 
operators with the correct quantum numbers, which have good overlap with the hadron 
wave function on the lattice) has to be constructed.

For that purpose we follow a strategy which already has been very successful in 
quenched simulations \cite{Burch:2006dg,Burch:2006cc}: We construct quark sources
of different spatial shapes and then construct a large number of interpolators from
these sources. As in the quenched case, by using two different gauge covariant 
smearings, we can mimic radial excitations. In addition we augment our basis with 
''p-wave sources'', i.e., sources which should have overlap with orbital excitations.

Previous work on excited and high spin mesons in the light quark sector can 
be found in Refs. \cite{Lacock:1996vy,Yamazaki:2001er,Petry:2008rt} for quenched calculations and in Refs. 
\cite{Aubin:2004wf,Allton:2004qq} for the dynamical case.

Preliminary results of this investigation were presented in \cite{Burch:2006rb,Burch:2007dh}. 
These studies are accompanied 
by similar ones on quenched lattices with chirally improved fermions \cite{Gattringer:2008be}.

In addition to these excitations in the low spin sector, we perform an exploratory
study of mesons with spin 2 and 3. 

In the following sections, we discuss the methods which we use to create the 
interpolators for our simulations and for extracting the excited states. Then, after 
briefly describing the details of the simulations, we present our findings for the 
meson spectrum. We give reasons for our choices of fit ranges and the expressions we 
have used for the chiral extrapolations. In the end we summarize our results for 
these channels and compare them to experimental values. All numerical details of our 
results are summarized at the end of this paper.

\section{The method}\label{SectMethod}

\subsection{Low spin}

Our calculation of the excited states of spin-0 and spin-1 mesons is based 
upon the variational method \cite{Michael:1985ne,Luscher:1990ck}. The idea is to use several 
different interpolators $O_i, i = 1, \ldots \, N$ with the quantum numbers 
of the desired state and to compute all cross correlations 
\begin{equation}
C(t)_{ij} \; = \; \langle \, O_i(t) \, \overline{O}_j(0) \, \rangle \; . 
\end{equation}
In Hilbert space these correlators have the decomposition 
\begin{equation}
C(t)_{ij} \; = \; \sum_n \langle \, 0 \, | \, O_i \, | \, n \, \rangle 
\langle \, n \, | \, O_j^\dagger \, | \, 0 \, \rangle \, e^{-t \, E_n}  \; . 
\label{corrmatrix}
\end{equation}
Using the factorization of the amplitudes one can show \cite{Luscher:1990ck} 
that the eigenvalues 
$\lambda^{(k)}(t)$ of the generalized eigenvalue problem 
\begin{equation}
C(t) \, \vec{v}^{(k)} \; \; = \; \; \lambda^{(k)}(t) \, C(t_0) \, 
\vec{v}^{(k)} \; , 
\label{generalized}
\end{equation}
behave as 
\begin{equation}
\lambda^{(k)}(t) \; \propto \; e^{-(t-t_0) \, M_k} \,[ \, 1 +
{\cal O}(e^{-(t-t_0) \, \Delta E_k}) \,] \; , 
\end{equation}
where $E_k = \sqrt{\vec{p}^2 +m^2}$ is the energy of the $k$-th state and 
$\Delta E_k$ is the difference to the energy closest to $E_k$ \cite{WeMo06}.
For a more detailed discussion of the error terms see \cite{Blossier:2008tx,Blossier:2009kd}.
In Eq.\ (\ref{generalized}) the eigenvalue problem is normalized with respect 
to a time slice $t_0 < t$.

In order to obtain mass spectra of states a spatial Fourier transformation is 
applied to the sink operator in order to project to zero momentum
\begin{equation}
O_i(t)\; = \; O_i(t,\vec{p}=0) \; = \; \frac1{V_3} \sum_{\vec{x}} O_i(t,\vec{x})
\, e^{i\vec{p}\cdot\vec{x}} \; ,
\end{equation}
with $\vec{p}=0$.

At this point we mention two other remarkable properties of the variational
method. First, it can be used to separate ghost contributions, as they appear 
in quenched or partially quenched calculations, from proper physical states.
In the spectral decomposition (\ref{corrmatrix}) ghosts appear with a modified 
time dependence. In \cite{Burch:2005wd} it was shown that the ghost 
contribution couples to an individual eigenvalue (up to the correction term).
Thus, these eigenvalues can be excluded from the analysis of the desired 
states. No modeling is necessary and thus no further uncertainties are 
introduced. 

Second, the eigenvectors of the generalized eigenvalue problem 
(\ref{generalized}) can be used to optimize the interval for fitting the
eigenvalues. If one plots the entries of the eigenvector $\vec{v}^{(k)}$ as a 
function of t, one finds that they form a plateau essentially in the same 
interval as the effective mass. Only in the time interval where both,
eigenvector components and effective mass, form a steady plateau, a fit to the
eigenvalues is unambiguous. Furthermore, the eigenvectors contain information
about the strength with which the different basis interpolators couple to a 
hadronic state. Thus, one can view them as a ``fingerprint'' of the corresponding 
state.

The variational method heavily relies on a basis of operator which has a large 
overlap with the states one is interested in. To construct such a basis we use 
several differently smeared quark sources. In a previous, quenched study 
\cite{Burch:2006dg,Burch:2006cc} we have optimized the 
smearings to match Gaussians which are ground and excited states of a 
spherical harmonic oscillator. However, a first study on dynamical 
clover-Wilson lattices has shown that it is very difficult to perform a 
similar matching. The reason is that with changing sea-quark mass the
lattice spacing also changes which means that the smearing parameters have to be
tuned for each set of configurations. In order to avoid such fine tuning 
procedures we simplify our construction of quark sources by using only a single
Gaussian source which we generate via Jacobi smearing \cite{Gusken:1989ad,Best:1997qp}.
The idea of Jacobi smearing is to create an extended source by 
iteratively applying the hopping part of the Wilson term (without Dirac 
structure) within the time slice of source and sink:
\begin{align}
b^{(\alpha,a)} &  =   S_J \; P^{(\alpha,a)} 
\; \; \; , \; \; \; \; 
S_J = \sum_{n=0}^N \, \kappa^n \, H^n \; , 
\nonumber \\
H(\vec{x},\vec{y}\,) &  =  \sum_{i = 1}^3
\Big[ \, U_i(\vec{x}) \, \delta_{\vec{x}+\hat{i},\vec{y}}
+ U_{-i}(\vec{x}) \, 
\delta_{\vec{x}-\hat{i},\vec{y}}
\, \Big] \, . 
\label{jacobismear}
\end{align}
We refer to the so constructed source as \textit{narrow source} in the following 
and denote it with $n$. In order to still allow for a radial excitation we also 
include a source where we apply a three-dimensional gauge covariant lattice 
Laplacian
\begin{align}
\Delta^{(3)}(\vec{x},\vec{y}\,) & = \sum_{i=1}^3 \left( U_i(\vec{x}) \, \delta_{\vec{x}+\hat{i},\vec{y}} + U_{-i}(\vec{x}) \, \delta_{\vec{x}-\hat{i},\vec{y}} - 2\delta_{\vec{x},\vec{y}} \right)
\label{laplacian3}
\end{align}
onto the narrow smeared sources. This one we call \textit{Laplacian source} and 
denote it with $L$. Since both Jacobi smearing and the Laplacian are scalar 
operators, these do not change the quantum numbers of our generic meson 
interpolators.

Further operators can be added to the operator basis by also exploring the 
possibility of orbital excitations. To do so we include in our quark sources 
additional \textit{derivative sources}. They are constructed by applying a 
symmetric covariant lattice derivative 
\begin{align}
\nabla_i(\vec{x},\vec{y}\,) & = \frac{1}{2} \left( U_i(\vec{x}) \, \delta_{\vec{x}+\hat{i},\vec{y}} - U_{-i}(\vec{x}) \, \delta_{\vec{x}-\hat{i},\vec{y}} \right)
\label{derivative}
\end{align}
in the appropriate direction onto the narrow smeared source. However, the 
resulting derivative sources, denoted by $\nabla_x$, $\nabla_y$, and 
$\nabla_z$, have to be combined appropriately with Dirac gamma matrices, to 
construct meson interpolators with the desired quantum numbers. The necessary 
group theory for this can be found in \cite{Liao:2002rj} and is later also 
used to construct operators for high spin mesons.

Finally, we also incorporate pointlike sources, denoted by $P$, to our set 
of smearings. Although the resulting interpolators have smaller overlap with 
the states, these additional sources give us the opportunity to not only 
extract the masses of the mesons, but also to compute local matrix elements 
which can be related to the decay constants of the mesons.

To summarize, we use the six different quark sources 
listed in Table \ref{quarksmearings}: a point source $P$, 
a narrow smeared source $n$, a source $L$, where a covariant spatial laplacian 
is applied to the narrow source, and derivative sources $\nabla_x$, $\nabla_y$, 
and $\nabla_z$. The latter ones are created by applying a covariant derivative 
in the corresponding spatial direction onto the narrow source. For the narrow 
source we use Jacobi smearing with fixed parameters $(N=8,\kappa=0.20)$.

\begin{table}
\begin{center}
\begin{tabular}{|c|l|}
\hline
$P$   & Point source at $x=0$ \\
$n$   & Narrow source from Jacobi smearing $P$ \\
$L$   & Covariant 3D lattice Laplacian applied on $n$ \\
$\nabla_x$ & Covariant derivative $\nabla_x$ applied on $n$ \\
$\nabla_y$ & Covariant derivative $\nabla_y$ applied on $n$ \\
$\nabla_z$ & Covariant derivative $\nabla_z$ applied on $n$ \\ \hline
\end{tabular}
\end{center}
\caption{List of the quark sources used and their specific smearing operations.}
\label{quarksmearings}
\end{table}

\subsection{High spin}

For the high spin mesons we try to extract only the ground states at the moment.
Therefore, we can restrict ourselves to single correlators.

The meson interpolators we use for this purpose are taken from the paper of X. 
Liao and T. Manke \cite{Liao:2002rj} which have been already used for calculating 
excited charmonium states \cite{Ehmann:2007hj}. These 
operators contain certain combinations of Dirac $\gamma$ matrices and necessarily 
also lattice derivative operators to be able to reach spin 2 and 3.

In discrete space-time one can only construct interpolators with definite lattice 
quantum numbers $R^{PC}$, in which $R$ is one of the five irreducible 
representations of the cubic group, namely $A_1, A_2, E, T_1$ and $T_2$. In order 
to determine the continuous quantum numbers $J^{PC}$ one has to map the finite 
number of irreducible representations of the cubic group to the infinite number of 
irreducible representations of the continuous rotation group. This is complicated 
by the fact that this projection is not unique. The mapping from $R$ to $J$ for 
the first lowest spin states is given by
\begin{align}
\label{mapRtoJ}
\begin{split}
A_1 & \rightarrow  J = 0,4,\ldots\\
A_2 & \rightarrow  J = 3,\ldots\\
T_1 & \rightarrow  J = 1,3,4,\ldots\\
T_2 & \rightarrow  J = 2,3,4,\ldots\\
E   & \rightarrow  J = 2,4,\ldots
\end{split}
\end{align}
Of course, for our simulations, we are especially interested in operators which 
transform according to $T_2$ and $E$, as well as $A_2$, since the lowest continuum 
spins to which these couple are $J=2$ and $J=3$, respectively.
To construct the interpolators for the high spin states we again combine one
or two gauge covariant lattice derivatives [see Eq. \ref{derivative}] with appropriate
Dirac gamma matrices according to Ref.~\cite{Ehmann:2007hj}. Also, here we want to 
improve the overlap with the physical states. Therefore, we apply to each quark source a gauge 
invariant Gaussian smearing using a spatial width of $2.4a$ and $16$ iterations. 
However, the correlators in these channels turn out to be particularly noisy. To
further improve our results we use APE smeared links ($\alpha=2.5$ and $N=15$), 
but only to create the source and the sink meson interpolators.

\section{Simulation details}

We calculate our meson correlators on configurations with two flavors of 
dynamical quarks. These configurations have been generated by the CP-PACS 
Collaboration using clover-Wilson fermions \cite{Sheikholeslami:1985ij} with a 
mean field improved clover coefficient and an RG improved gauge action 
\cite{Iwasaki:1985we}.

In Table \ref{lattices}, we summarize details of the configurations used in our 
simulations. For each lattice size, there exist four ensembles with different 
sea-quark mass. The values $\kappa_{sea}$ have been chosen in such a way that the 
ratio $m_{PS}/m_V$ is approximately the same for the different lattice sizes. 
For our simulations we use only every fifth available configuration in each ensemble
in order to reduce effects coming from finite autocorrelation times and at the
same time save computational resources. Thus, we analyze between 80 and 150 
configurations per ensemble for our calculations (see Table \ref{lattices}).
There exists also an even finer lattice with $24^3\times 48$ sites, but we lacked 
the computer time to make use of it. More information 
about these configurations, especially on how they have been generated can be found 
in Refs.~\cite{Aoki:1999ff,Ali Khan:2001tx}.

\begin{table}
\begin{center}
\begin{tabular}{|c|c|c|c|c|c|c|}
\hline
$\beta$ & $L^3 \times T$ & $c_{SW}$ & $a[fm]$      & $La[fm]$    & $\kappa_{sea}$ & \#configs \\  \hline
$1.80$  & $12^3\times24$ & $1.60$   & $0.2150(22)$ & $2.580(26)$ & 0.1409         & 129       \\
        &                &          &              &             & 0.1430         & 104       \\
        &                &          &              &             & 0.1445         & 144       \\
        &                &          &              &             & 0.1464         & 80        \\  \hline
$1.95$  & $16^3\times32$ & $1.53$   & $0.1555(17)$ & $2.488(27)$ & 0.1375         & 113       \\
        &                &          &              &             & 0.1390         & 137       \\
        &                &          &              &             & 0.1400         & 137       \\
        &                &          &              &             & 0.1410         & 99        \\  \hline
\end{tabular}
\end{center}
\caption{Details about the CP-PACS configurations used. The values are taken from Refs. 
\cite{Aoki:1999ff,Ali Khan:2001tx}.}
\label{lattices}
\end{table}

For our simulations we used Chroma \cite{Edwards:2004sx}. At the time this project 
started only version 2.2.1 of this lattice QCD library was available, which did 
not contain a proper implementation of clover-Wilson fermions, Jacobi smearing, 
and the calculation of cross-correlation matrices. Therefore, we developed our own 
routines for these tasks. Starting with version 3 of Chroma, the above-mentioned 
routines were also implemented. For consistency, we stick to our own routines for 
calculating excited states in the low spin sector, while using native Chroma with 
appropriate XML input files for the high spin mesons. Thus, for the high spin 
sector, we can take advantage of different optimizations, like the SSE optimized 
Wilson Dslash \cite{chroma1} and Peter Boyle's BAGEL \cite{chroma2} for running on 
QCDOC \cite{Boyle:2003mj}.

\section{Results}

\subsection{Low spin}
\label{lowspin}
\subsubsection{Effective masses}

In the following we present the results of our calculations. For our analysis we 
take advantage of several symmetries of the cross-correlation matrix. We find that 
the matrices $C(t)$ are real and symmetric within error bars. Therefore, we 
symmetrize them by replacing $C_{ij}(t)$ with $(C_{ij}(t)+C_{ji}(t))/2$. We can 
increase our statistics even further by taking into account the contributions which 
are proportional to $\exp[-(T-t)M_n]$. We symmetrize our correlators by replacing 
$C(t)$ with $( |C(t)| + | C(T-t) | )/2$ and use the resulting matrix in the variational 
method.

The eigenvalues, which we obtain from the generalized eigenvalue problem (\ref{generalized}), 
can then be fitted to the function
\begin{eqnarray}
\label{evals}
\lambda^{(k)} (t,t_0) & = & Ae^{-M_k(t-t_0)},
\end{eqnarray}
where we use $t_0=1$ in all cases. Ideally one should use the largest possible value of $t_0$;
however, due to the limited signal for the excited states, we are not able to go to larger values.
To determine fit ranges, we define the effective mass
\begin{eqnarray}
m_{eff}(t+\frac{1}{2}) & = & \ln \left( \frac{\lambda^{(k)}(t,t_0)}{\lambda^{(k)}(t+1,t_0)} \right).
\end{eqnarray}
This quantity should form a plateau as a function of $t$ once the contributions of 
the higher excited states are strongly suppressed. Additional information is provided 
by the eigenvectors $\vec{v}^{(k)}$. Their components should also show a plateau when 
only a single state contributes.

Another important feature of our analysis is that we use only a submatrix of the 
correlator matrix. We refer to this procedure as \textit{pruning of the operator 
basis}. The reason is that many of the interpolators have only small overlap with the 
physical state or, given the limited number of gauge configurations, they convey no 
new information. Their inclusion contributes mainly noise to the correlator. Also, 
one can show \cite{Allton:1993wc} that choosing certain interpolator combinations 
helps in suppressing contributions of higher order corrections in the different 
eigenvalues. In this way, one can improve the effective mass plateaus to a certain 
extent by choosing an optimal, and often smaller, interpolator basis. However, to 
find such a combination is rather difficult, since the number of possibilities to 
choose a certain interpolator combination is extremely high.
Of course we cannot check all possible combinations; however, one quickly learns which 
operators are worth leaving out. This might also make pruning 
very subjective and thus can lead to ambiguous results if several combinations of 
operators seem to be equally good but give slightly different effective mass plateaus. 
As long as these deviations are well within errors we should be allowed to choose 
anyone of these combinations. We choose those that appear to give the best effective mass
plateaus from the largest possible number of eigenvalues (here, 2 for the pseudoscalars 
and vectors and 1 for the rest).

\begin{figure*}
\includegraphics*[width=0.6\textwidth,clip]{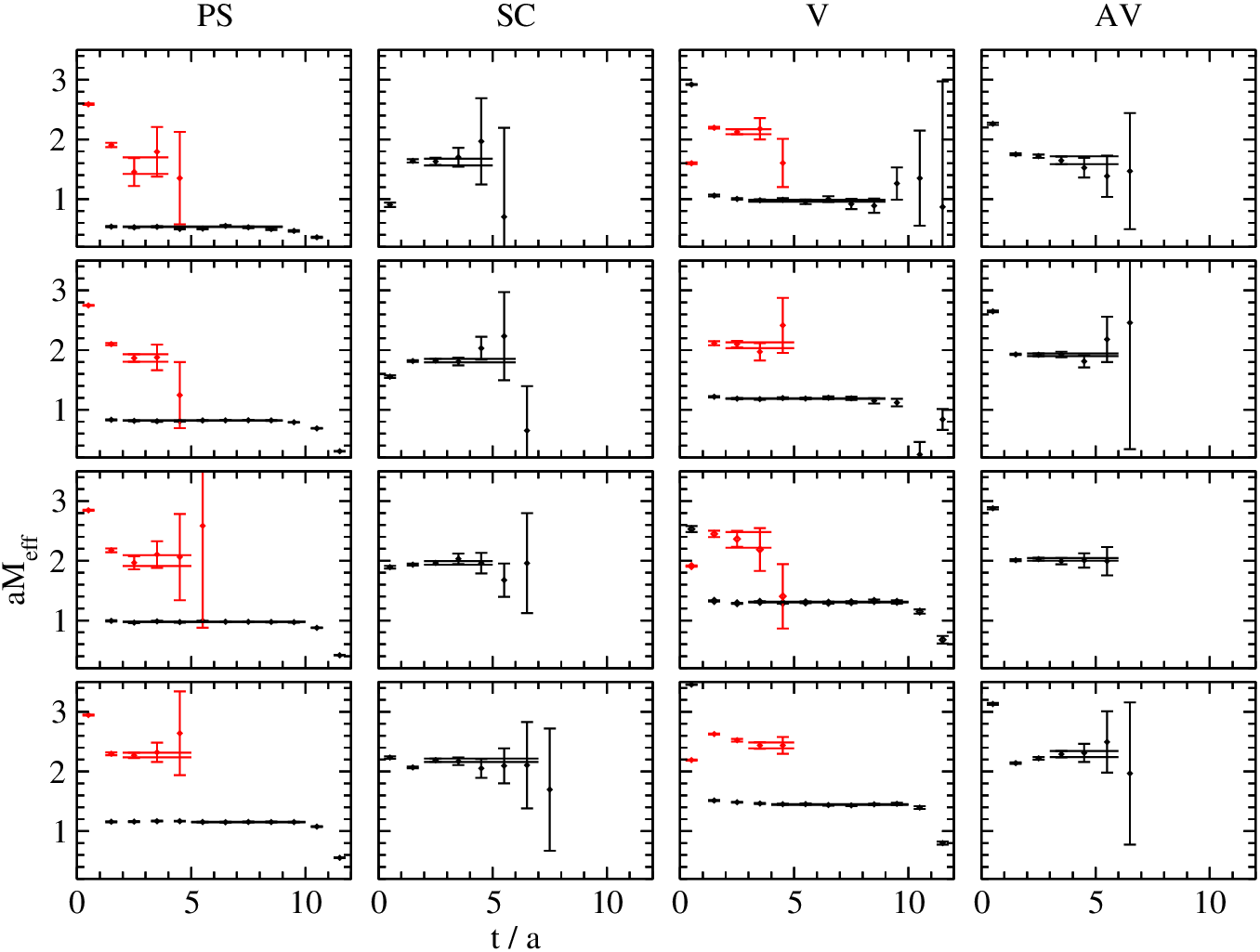}
\caption{Effective mass plots for mesons from our coarse lattice (with 
$\kappa=0.1464,0.1445,0.1430,0.1409$ from top to bottom). Both ground and excited states are 
shown, along with the $M\pm\sigma_M^{}$ results (horizontal lines) from correlated fits to 
the corresponding time intervals. For the PS channel, we show results for both operator 
combinations.}
\label{effmass12}
\end{figure*}

In Fig. \ref{effmass12}, we show the effective masses for pseudoscalar (PS), scalar 
(SC), vector (V), and axialvector (AV) mesons obtained on the $12^3\times 24$ lattice 
for the four quark masses we have used in our calculations. The horizontal lines 
denote the time intervals where we have performed correlated fits to the eigenvalues 
and represent the resulting masses and their statistical errors.

We obtain excellent plateaus for the pseudoscalar and vector ground states. For these 
channels we are also able to extract first excited states. There, however, the results 
are not that good: The plateaus consist of only two or three effective mass points and 
are very noisy. We find that the ground states for both meson channels are practically 
unaffected by the choice of operators. For the excited pseudoscalar meson we are able 
to use the same optimal interpolator combination for all quark masses. However, to 
obtain results for the excited vector meson state we have to alter the optimal operator 
combination for each sea-quark mass (see Table \ref{tab:lowspinmasses12}).

The results for scalar and axialvector are also very good, however, slightly noisier 
than those of pseudoscalar and vector ground states. The fact that the pseudoscalar 
and vector channels yield better results than the other mesons is usually observed in 
lattice QCD. This is not unexpected since these states are much lighter than all the 
others and thus yield a better signal for a larger number of time slices.

\begin{figure*}
\includegraphics*[width=0.6\textwidth,clip]{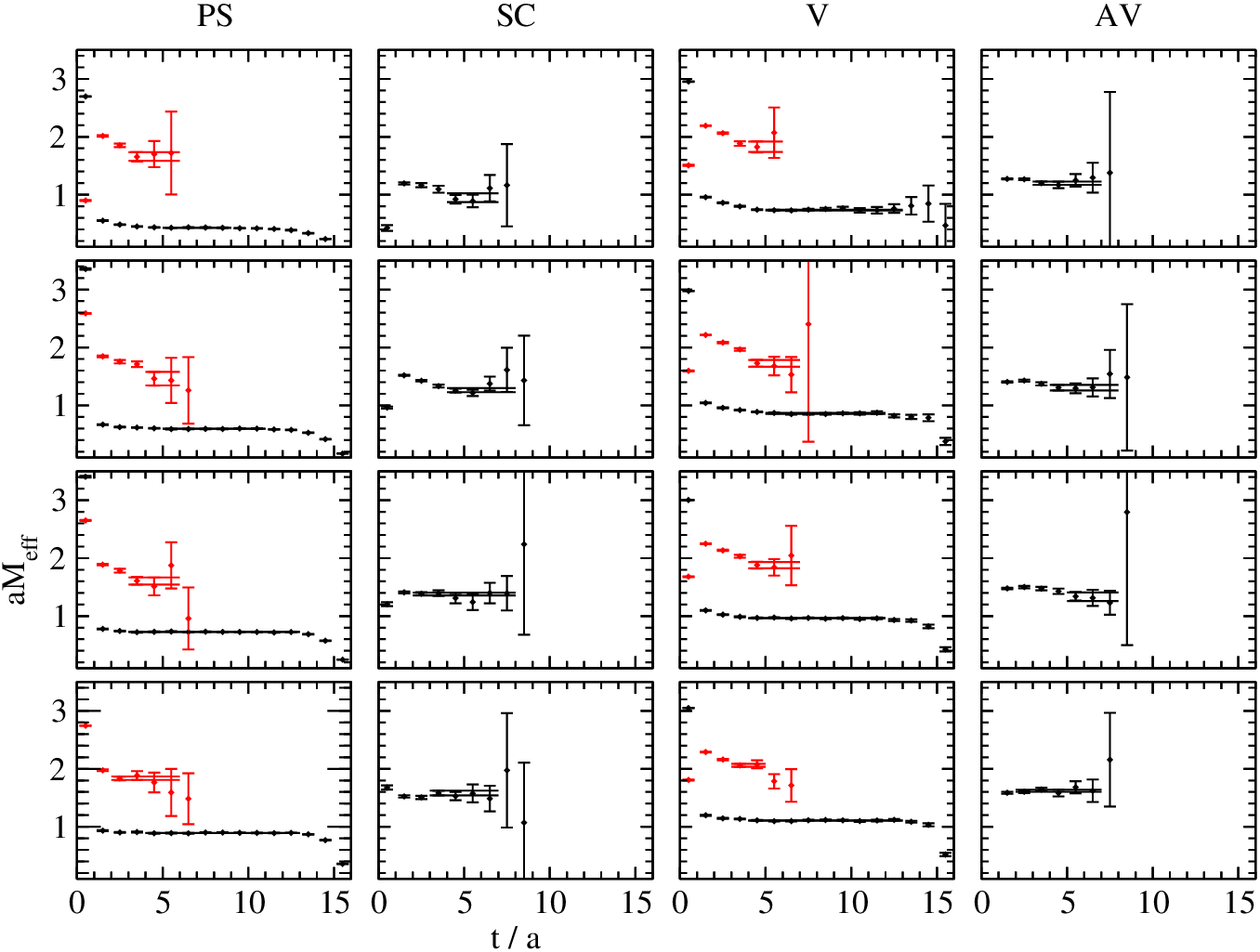}
\caption{Effective mass plots for mesons from our fine lattice (with 
$\kappa=0.1410,0.1400,0.1390,0.1375$ from top to bottom). Both ground and excited states are 
shown, along with the $M\pm\sigma_M^{}$ results (horizontal lines) from correlated fits to 
the corresponding time intervals. For the V channel, we show results for both operator 
combinations.}
\label{effmass16}
\end{figure*}

In Fig. \ref{effmass16}, we present the effective masses from the finer lattice.

Again we obtain excellent results for pseudoscalar and vector ground states with long 
clear plateaus. However, the situation for the excited pseudoscalar and vector states 
improved only marginally. The plateaus are noisy and rather short; often we can include 
only three or four time slices in our fits. Certainly an improvement is given by the fact 
that for the finer lattice we can choose the same optimal combination for all sea-quark 
masses, except for the smallest quark mass. There we altered the optimal interpolator 
combination for the pseudoscalar meson slightly (see Table \ref{tab:lowspinmasses16}).

In the scalar and axialvector channel we find only a slight improvement when going to 
the finer lattice. For the scalar meson it is necessary to choose a different operator 
for $\kappa=0.1400$ than for the other masses. Since the combination $L\mathds{1}n$ is 
very similar to $\nabla_i\mathds{1}\nabla_i$ (both of them represent a three-dimensional 
lattice Laplacian but with different displacement), we do not regard this as a problem.

Fortunately, in our previous quenched studies \cite{Burch:2006dg,Burch:2006cc}, we were 
able to use for each valence quark mass the same time slice as the starting point of the fit 
intervals. In this study, however, we sometimes need to change this time slice as we move 
from one quark mass to next one. The reason is that the ensembles for different sea-quark 
masses are generated independently. Thus, they should be completely uncorrelated, in 
contrast to the quenched case, where we changed only valence quark mass but always used 
the same set of configurations. Additionally, the effective lattice spacing depends on 
the sea-quark mass.  Nevertheless, we still require that both the effective mass and 
components of the corresponding eigenvector show plateaus in the fit interval.

The numerical results of our correlated fits together with the optimal operators for the meson states 
can be found in Tables \ref{tab:lowspinmasses12} and \ref{tab:lowspinmasses16}.

\subsubsection{Pseudoscalar meson ground state}

\begin{figure*}
\includegraphics*[width=0.4\textwidth,clip]{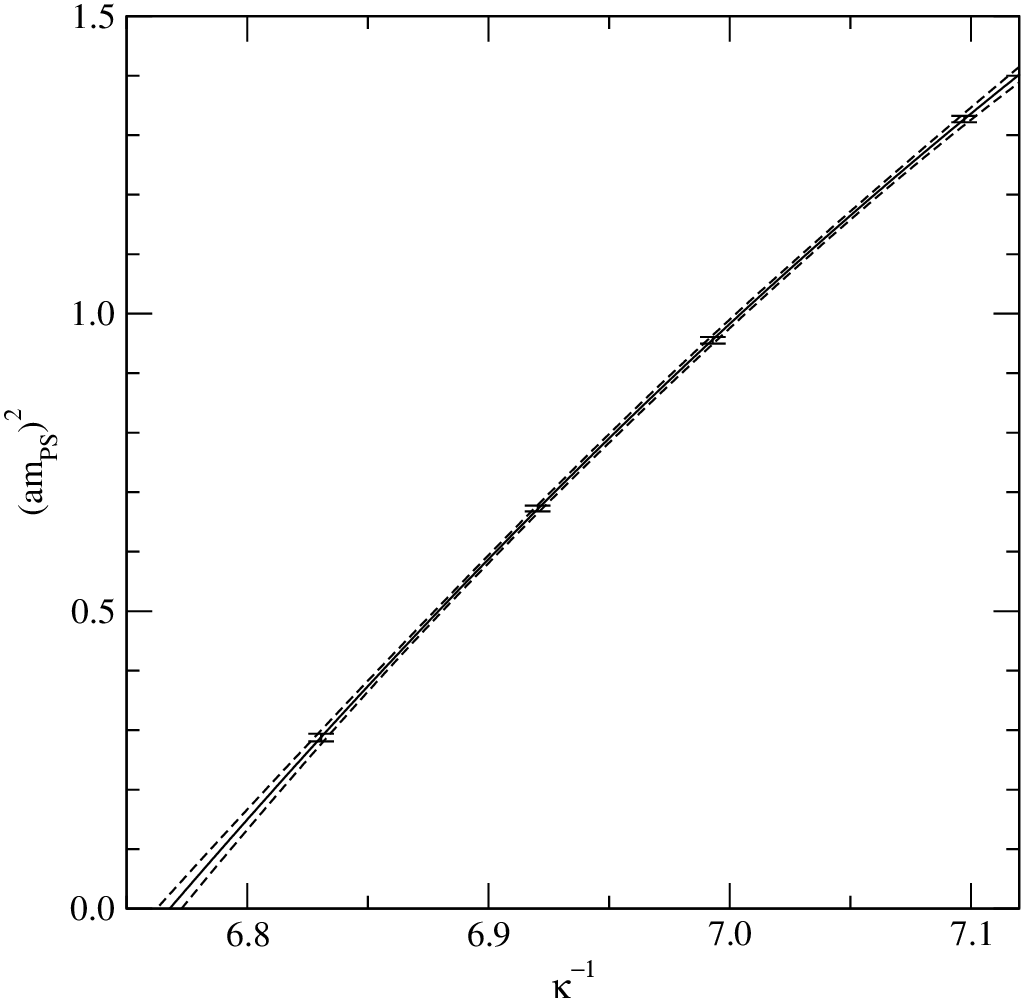}
\hspace*{5mm}
\includegraphics*[width=0.4\textwidth,clip]{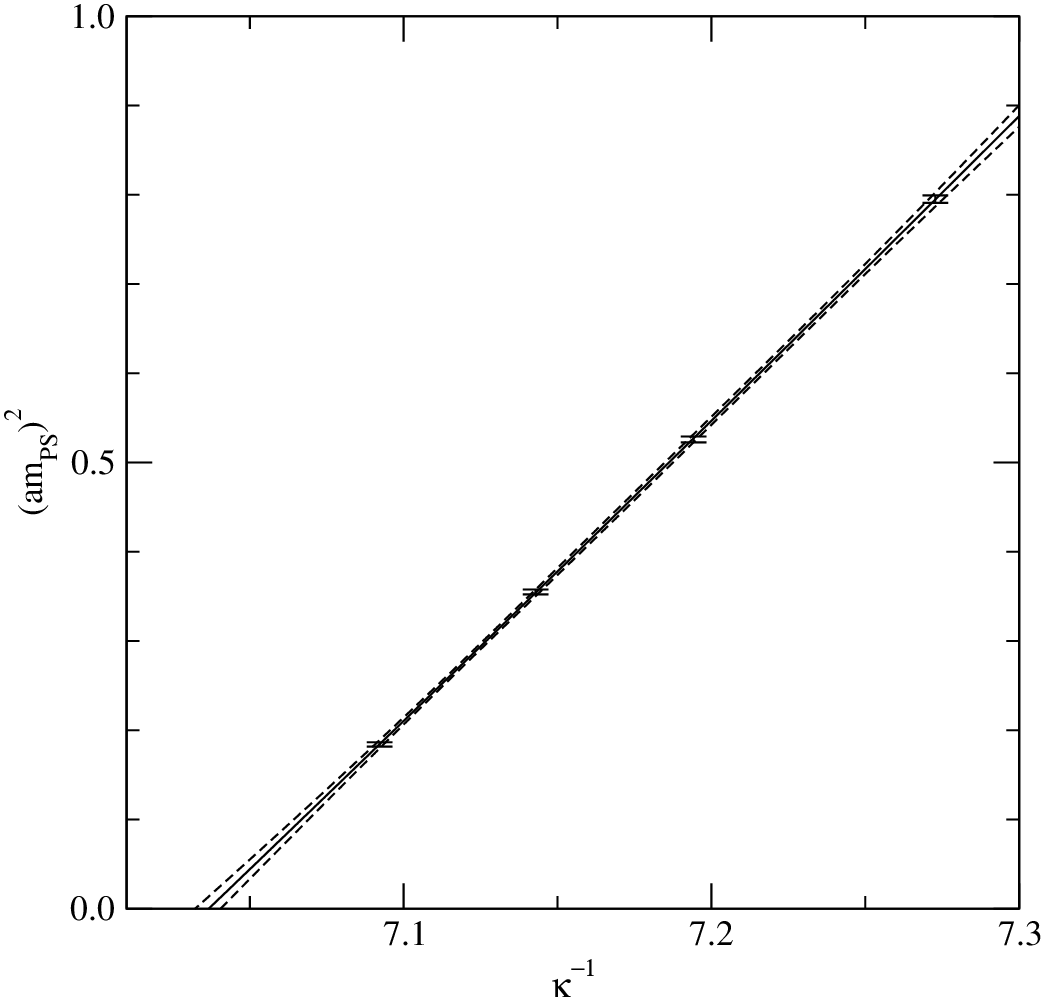}
\caption{The figure shows $(am_\pi)^2$ as a function of $\kappa^{-1}$. The left plot is for 
the $12^3\times24$ lattice, while the right plots shows the results for the $16^3\times32$ lattice.}
\label{fig:psground}
\end{figure*}

For the pion ground state the results of our fits are presented in Fig. \ref{fig:psground}, 
where we plot the pion mass squared as a function of $\kappa^{-1}$. To be able to 
extrapolate our other results to the chiral limit, we have to determine the critical 
quark mass. It is defined as the value $\kappa_c^{-1}$ where the mass of the pseudoscalar 
meson vanishes.

For the pseudoscalar meson the appropriate chiral extrapolation formula is given by the 
resummed Wilson chiral perturbation theory (RW$\chi$PT) \cite{Aoki:1999ff}. It reads
\begin{align}
m_{PS}^2 & = Am \left[ -\log\left( \frac{Am}{\Lambda_0^2} \right) \right]^{\omega_0} \left[ 1 + \omega_1 m \log \left(  \frac{Am}{\Lambda_3^2} \right) \right],
\end{align}
where $m=\frac{1}{2}(\kappa^{-1} - \kappa_c^{-1})$ is the quark mass and $A$, $\Lambda_0$, 
$\Lambda_3$, $\omega_0$, and $\omega_1$ are parameters in the theory. Since we have only 
four data points for each lattice it is not possible to use this expression as a fit 
function. Therefore, we restrict ourselves to a much simpler function given by
\begin{align}
 (am_{PS})^2 = B_{PS}\,m + C_{PS}\,m^2,
\end{align}
and we take $\kappa_c^{-1}$ as an additional fit parameter. The linear term is motivated 
by Wilson $\chi$PT without resummation
\begin{align}
 m_{PS}^2 & = Am \left[ 1 + \omega_1 m \log \left( \frac{Am}{\Lambda_3^2} \right) + \omega_0 \log \left( \frac{Am}{\Lambda_0^2} \right) \right],
\end{align}
while we include the quadratic term in order to account for the slight curvature of our 
results. Since we are working at pion masses from approximately $500$ MeV to 1 GeV, it is 
highly questionable to what extent $\chi$PT is applicable.

\subsubsection{Vector meson ground state}

In the upper two plots of Fig. \ref{fig:ground}, we present our results for the vector 
meson ground state as a function of the mass of the pseudoscalar ground state.

For the chiral extrapolations we use 
\begin{align}
 am_V = A_V + B_V (am_{PS})^2 + C_V (am_{PS})^4
\end{align}
as the fit function. 

Our results for the pion and rho ground states are slightly different from the ones obtained 
by the CP-PACS Collaboration. For consistency, we thus redetermine the physical point and 
the lattice spacing by following the procedure described in \cite{Ali Khan:2001tx}. For the 
physical point, we consider the ratio 
\begin{align}
\label{ppoint}
 \frac{am_\pi}{A_V + B_V (am_\pi)^2 + C_V (am_\pi)^4} = \frac{M_\pi}{M_\rho},
\end{align}
where $M_\pi = 0.1396$ GeV and $M_\rho = 0.7755$ GeV are fixed to the experimental values. 
The lattice spacing is then given by 
\begin{align}
 a = \frac{am_\rho}{M_\rho},
\end{align}
with $am_\rho = am_V(am_\pi)$ being the mass of the rho meson in lattice units determined at 
the physical point for $am_\pi$, determined via Eq.~(\ref{ppoint}). In addition, we can 
also compute $\kappa_{ud}^{-1}$ which corresponds to up/down quark mass on the lattices by 
solving
\begin{align}
 (am_{PS})^2(\kappa_{ud}) = (am_\pi)^2.
\end{align}

The resulting values for the physical point $am_\pi$, the lattice spacing $a$, and the 
parameters $am_\pi$ and $am_\rho$ are summarized in Table \ref{tab:physpoint}.

\subsubsection{Scalar and axialvector meson ground state}

After determining the physical point and the lattice spacing, we can discuss the results for 
the other meson channels.

\begin{figure*}
\includegraphics*[width=0.39\textwidth,clip]{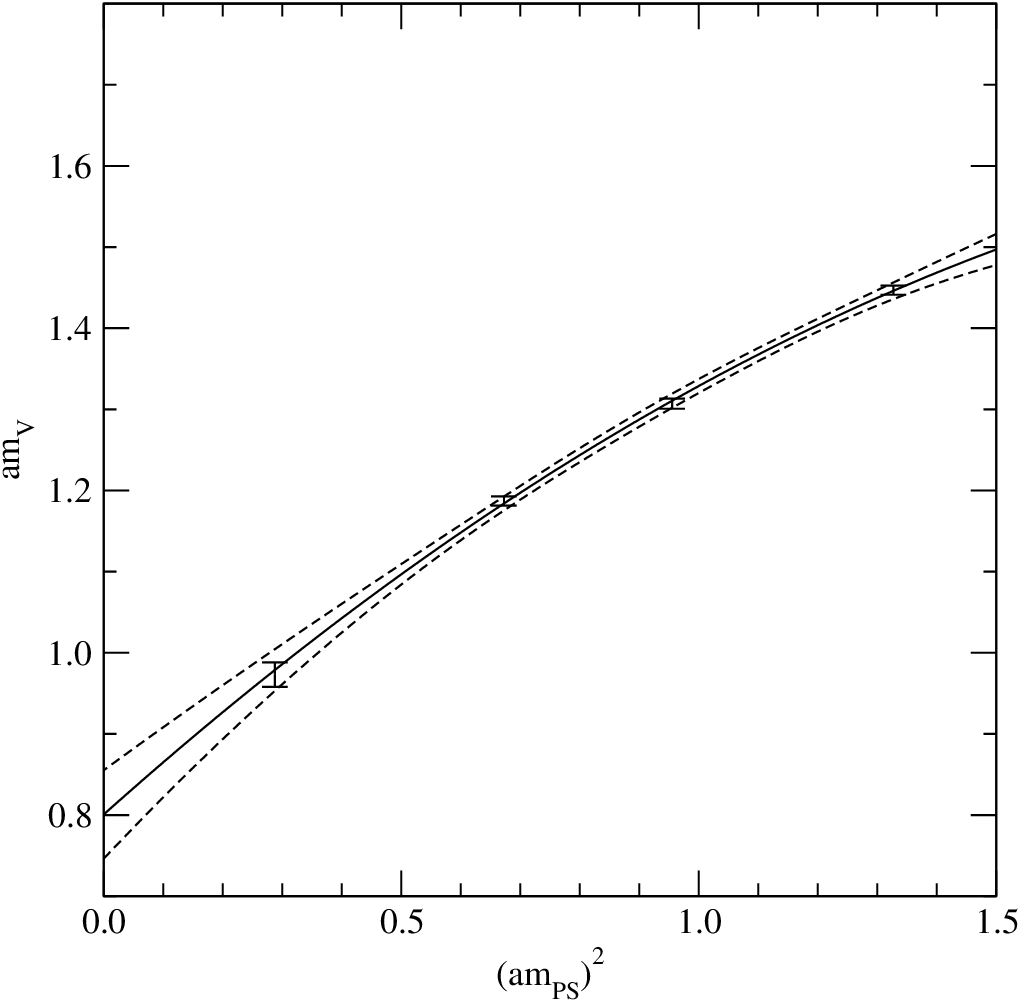}
\hspace*{5mm}
\includegraphics*[width=0.39\textwidth,clip]{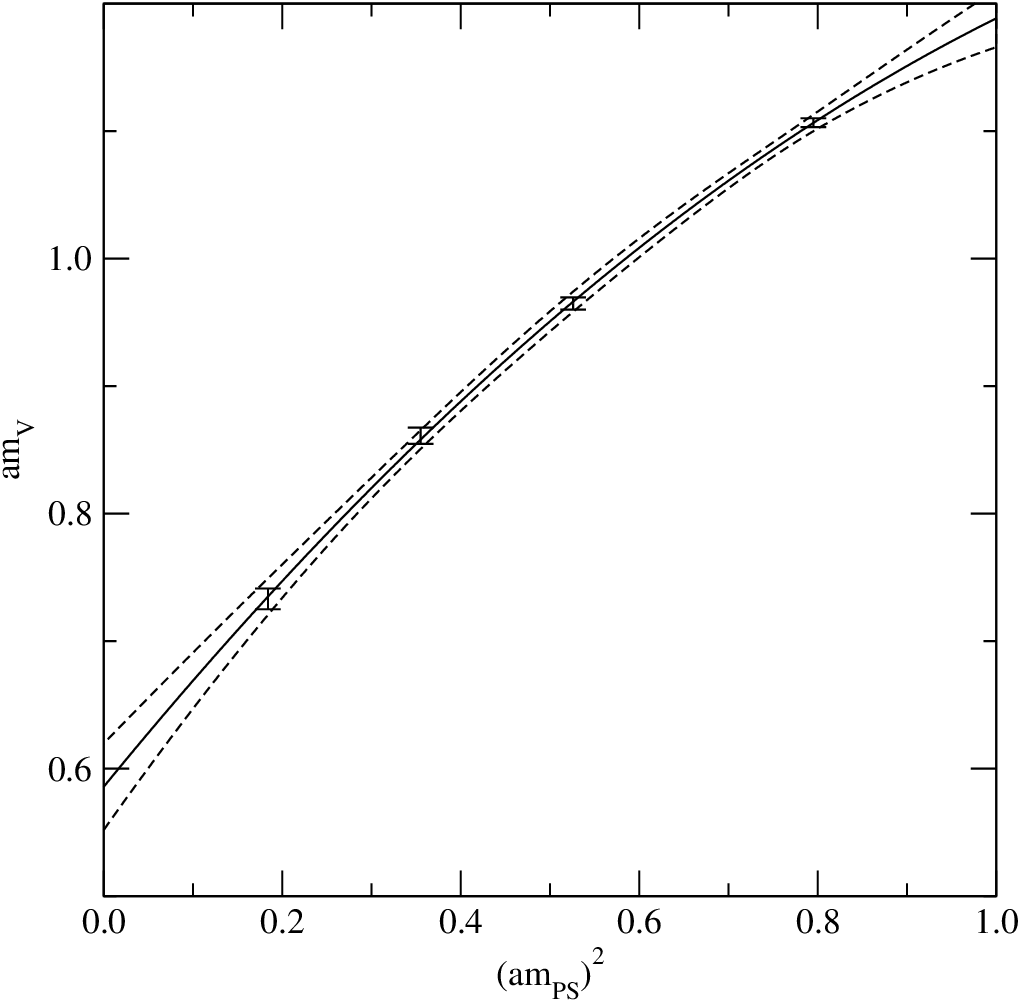}

\vspace*{5mm}

\includegraphics*[width=0.39\textwidth,clip]{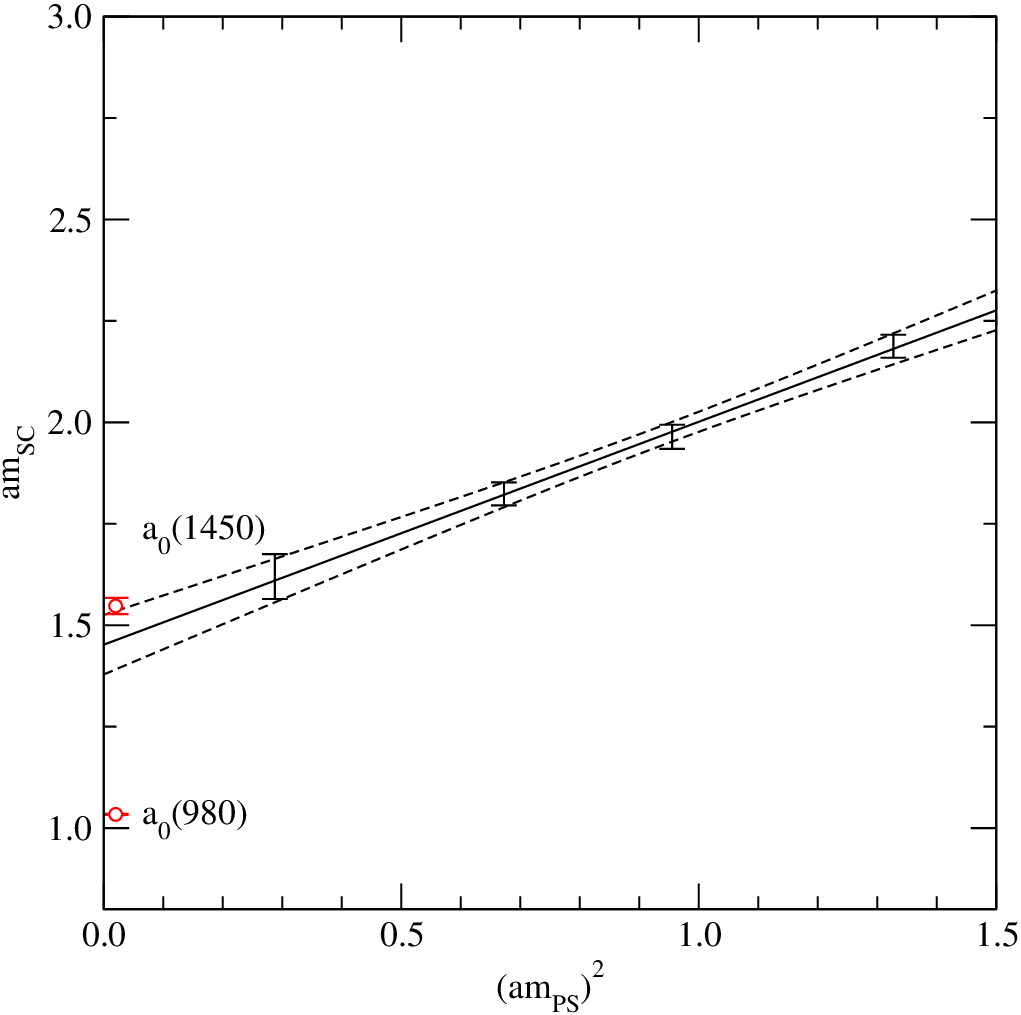}
\hspace*{5mm}
\includegraphics*[width=0.39\textwidth,clip]{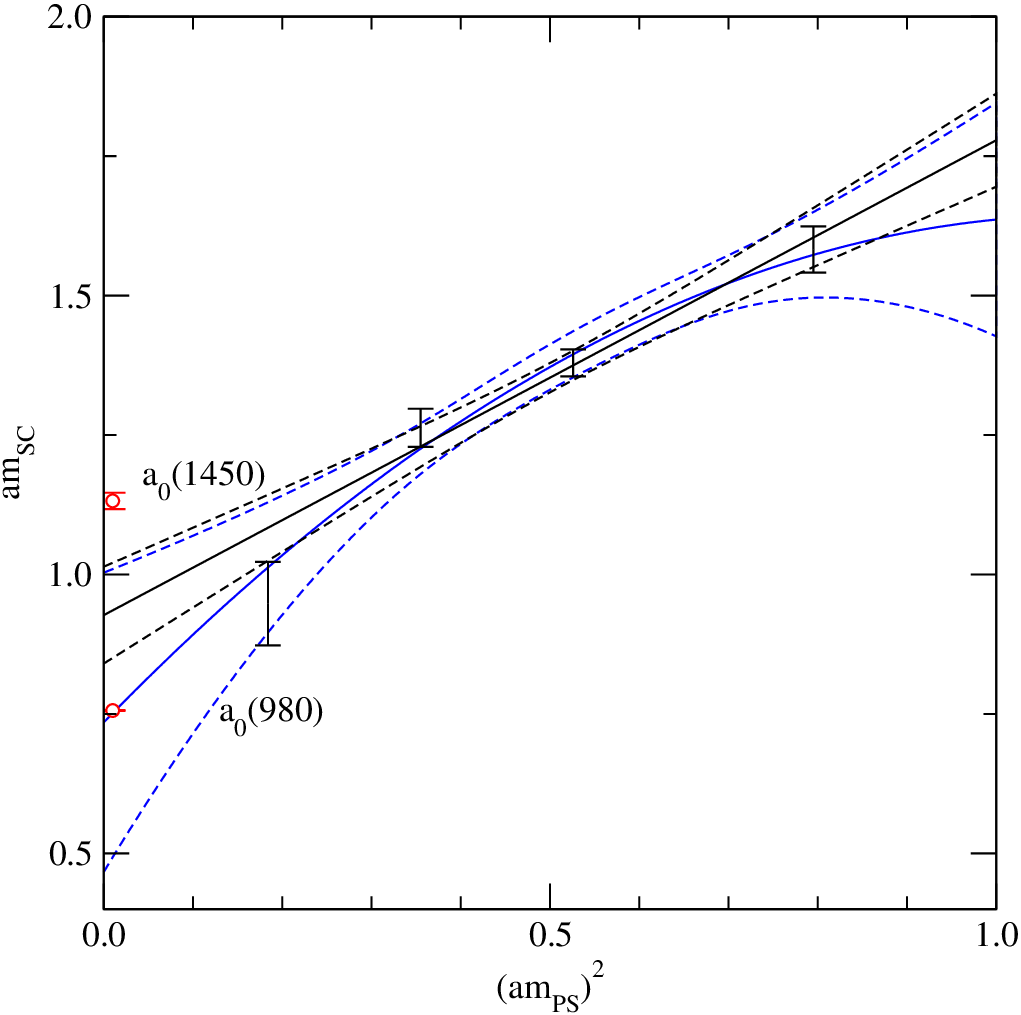}

\vspace*{5mm}

\includegraphics*[width=0.39\textwidth,clip]{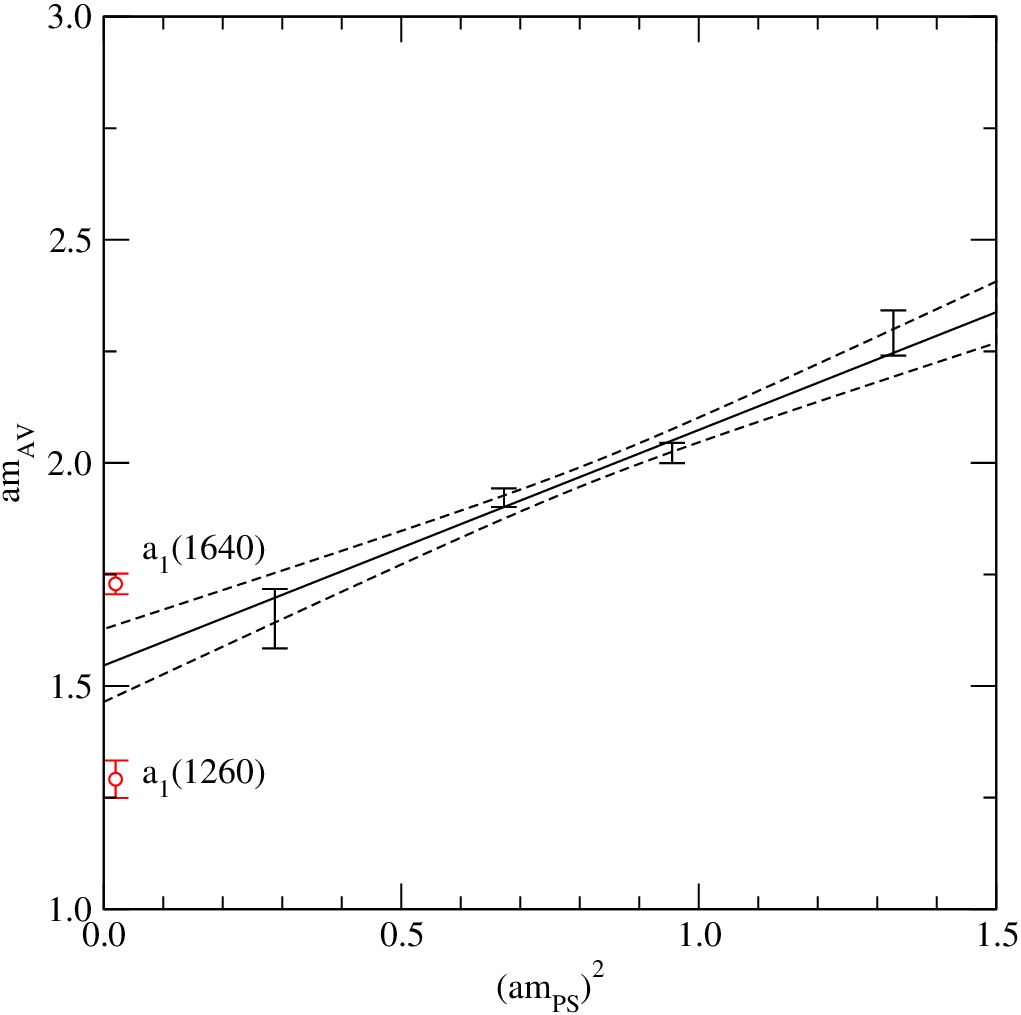}
\hspace*{5mm}
\includegraphics*[width=0.39\textwidth,clip]{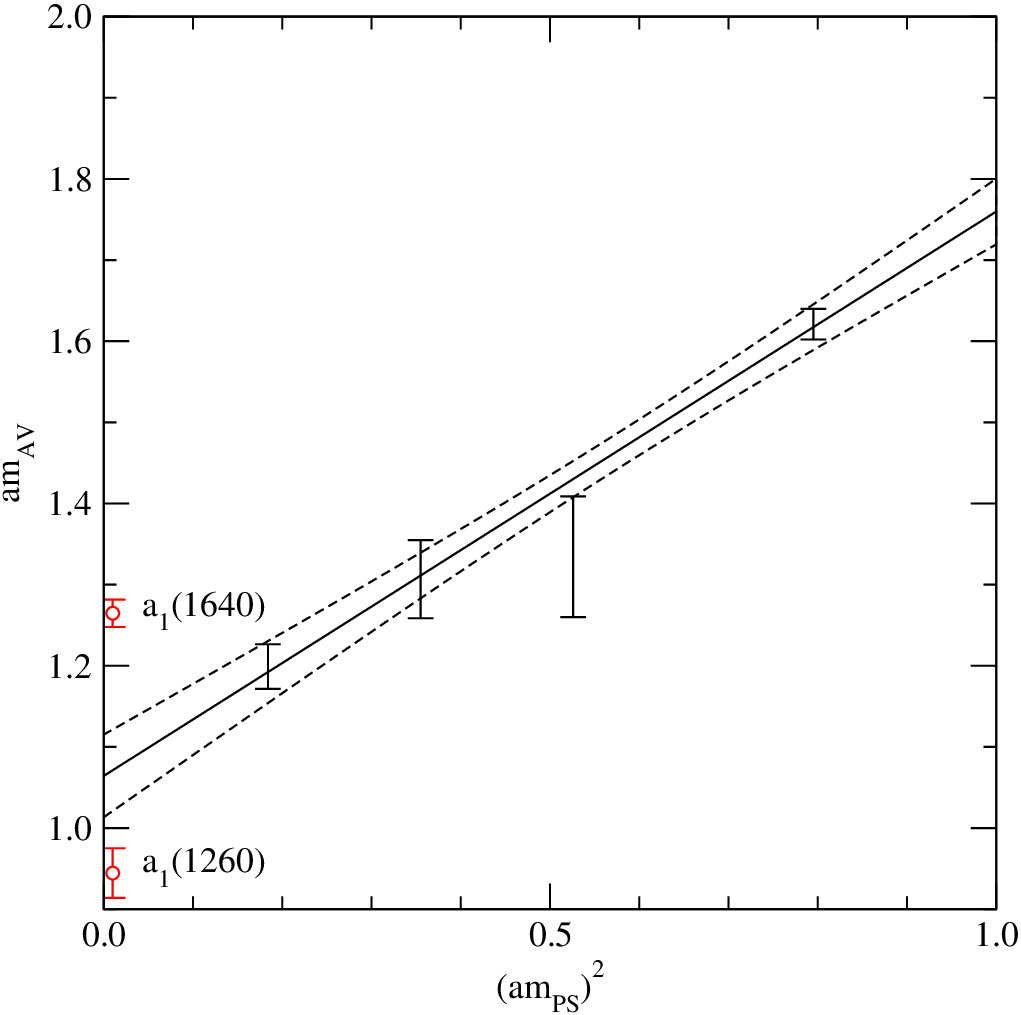}
\caption{The figure shows the vector, scalar, and axialvector meson ground states as a function 
of $(am_\pi)^2$. The left plots are for the $12^3\times24$ lattice, while the right plots show 
the results for the $16^3\times32$ lattice. The circles represent the experimental points. They 
are omitted for the vector meson since it is used to set the scale. We also show the results of 
our chiral extrapolation (solid line) together with the one sigma error band (dashed lines). For 
the scalar meson results on the $16^3\times32$ lattice both linear and quadratic fits have been 
performed.}
\label{fig:ground}
\end{figure*}

We start with the scalar ground state which is shown in the second row of plots in Fig. 
\ref{fig:ground}. For the $12^3\times24$ lattice, we find that the scalar mass depends linearly 
on the squared pion mass. Therefore, we perform linear fits in $(am_{PS})^2$ for the chiral 
extrapolation. This means that we fit our results to
\begin{align}
am_{SC} & = A_{SC} + B_{SC} (am_{PS})^2.
\label{scgrlin}
\end{align}
However, for the finer lattice the scalar meson mass for the smallest quark mass shows some 
deviation from the linear behavior of the other points. Therefore, we extend the expression 
in Eq.~(\ref{scgrlin}) by an additional term $ C_{SC} (am_{PS})^4$. We also try to add 
such an additional term to the fit functions of the other meson states. However, in all these 
cases the fit results for the corresponding parameter $C$ is consistent with zero.

For the axialvector meson ground state (see Fig. \ref{fig:ground} lower plots) we find that 
the results on both lattices depend linearly on $(am_{PS})^2$ . Thus, we use 
\begin{align}
am_{AV} & = A_{AV} + B_{AV} (am_{PS})^2
\end{align}
as the fit function for our chiral extrapolations. The only point which shows a slight deviation 
from a linear behavior is the point at $(am_{PS})^2 \approx 0.53$. Nevertheless, we have decided 
to include this point in our fit, since leaving it out changes our results negligibly.

\subsubsection{Pseudoscalar and vector meson excited state}

\begin{figure*}
\includegraphics*[width=0.4\textwidth,clip]{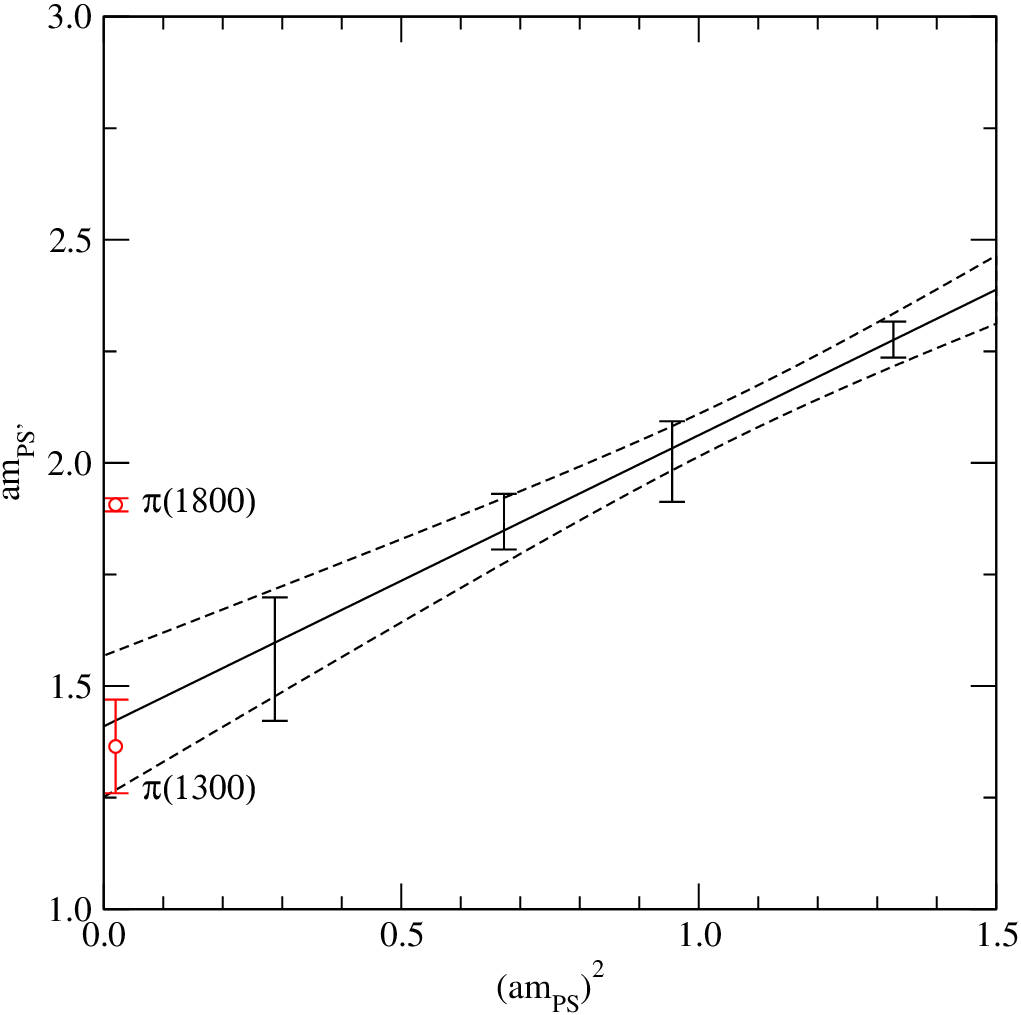}
\hspace*{5mm}
\includegraphics*[width=0.4\textwidth,clip]{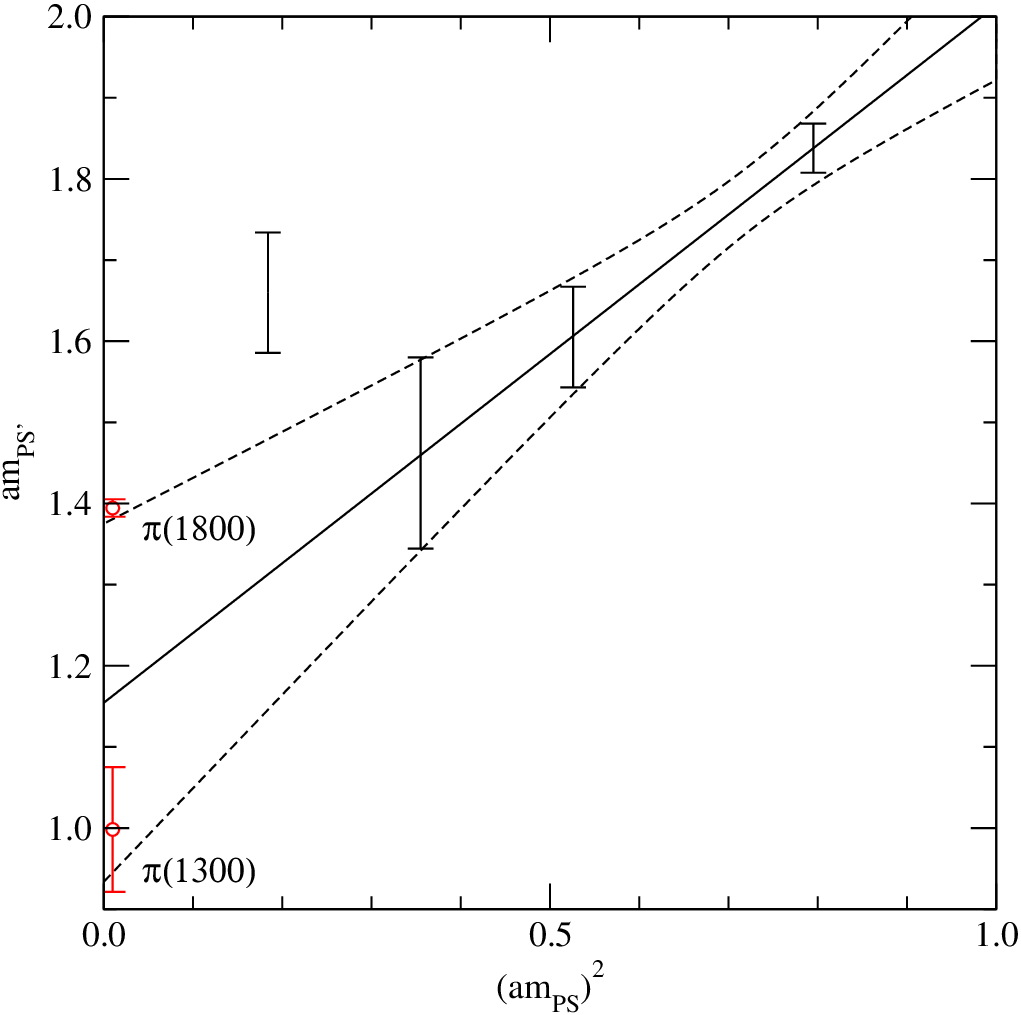}

\vspace*{5mm}

\includegraphics*[width=0.4\textwidth,clip]{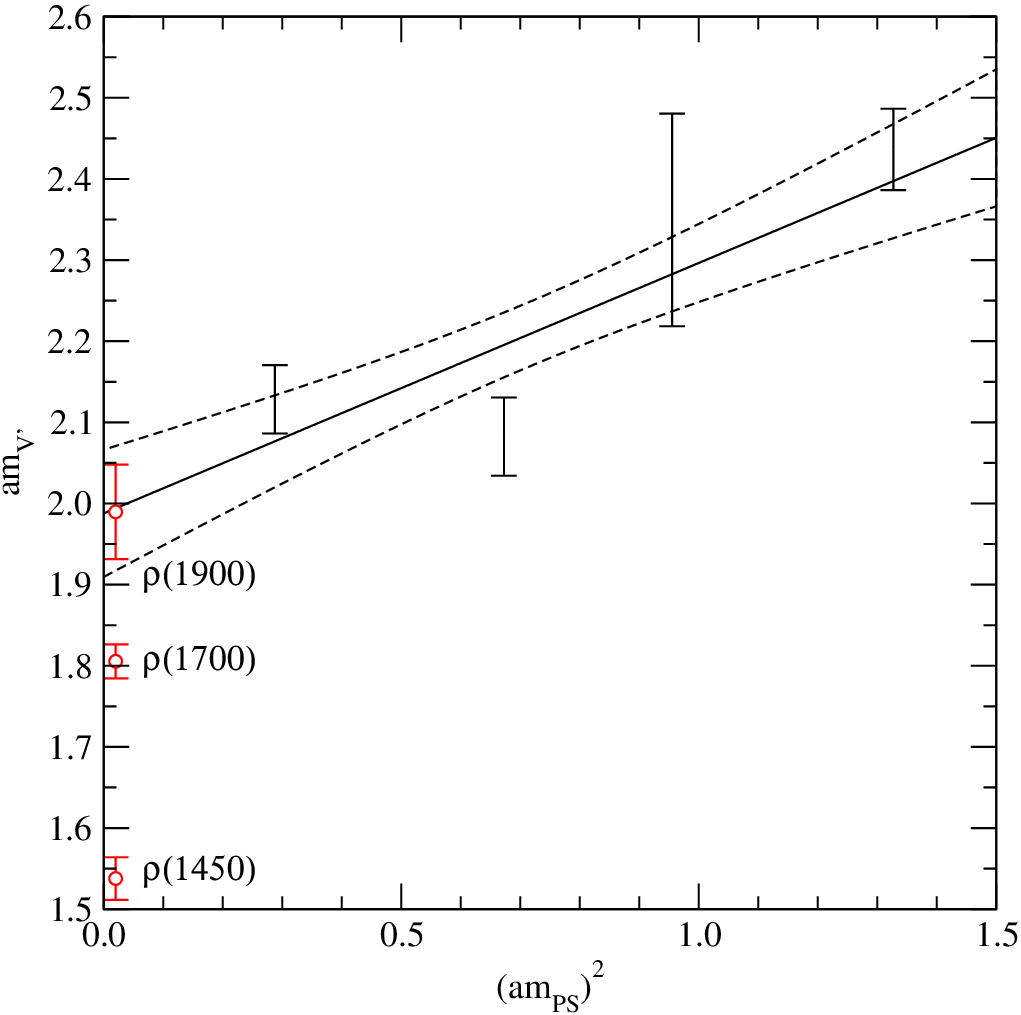}
\hspace*{5mm}
\includegraphics*[width=0.4\textwidth,clip]{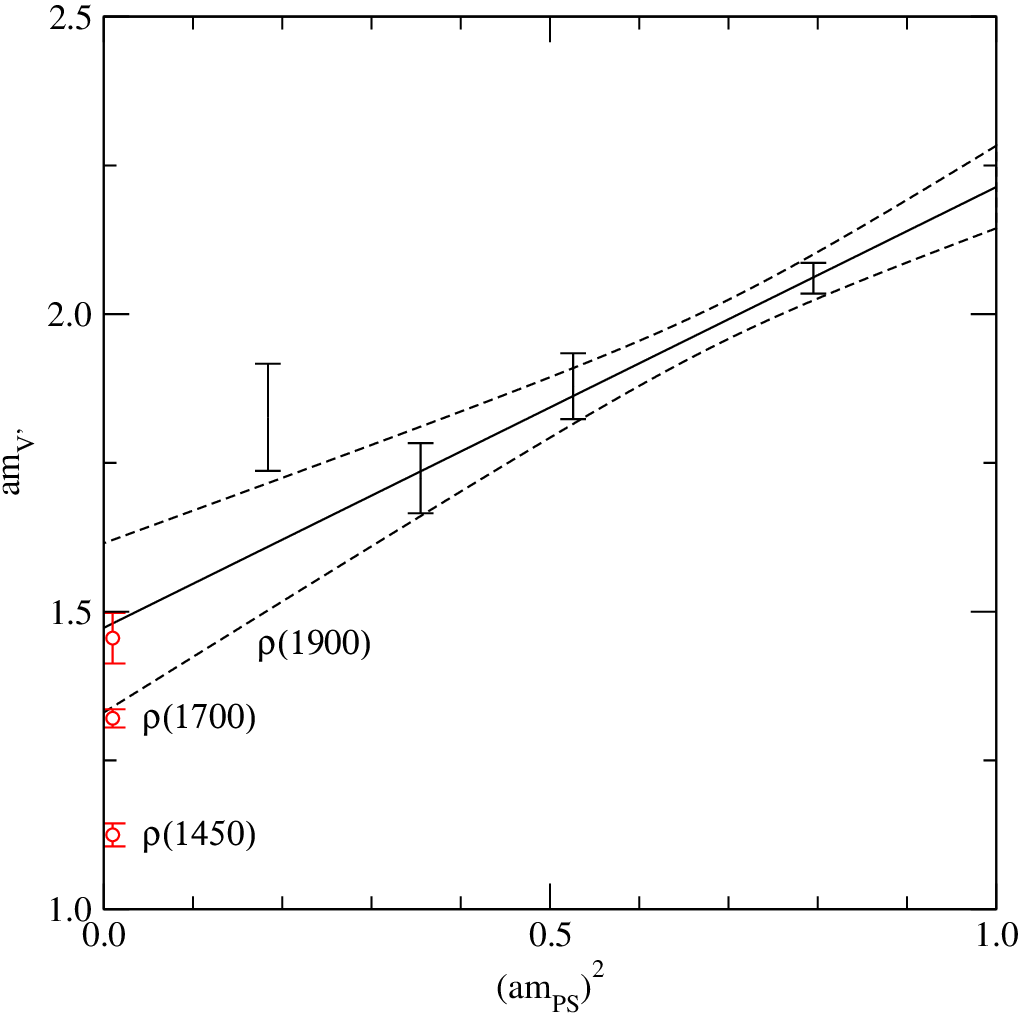}
\caption{The figure shows the pseudoscalar and vector meson first excited state as a function 
of $(am_\pi)^2$. The left plot is for the $12^3\times24$ lattice, while the right plot shows the 
results for the $16^3\times32$ lattice. The circles represent the experimental points. We also 
show the results of our chiral extrapolation (solid line) together with the one sigma error band 
(dashed lines).}
\label{fig:excited}
\end{figure*}

We start our discussion of the excited states with the excited pseudoscalar meson. In the upper 
plots in Fig. \ref{fig:excited} we plot the results of our fits to the eigenvalues as a 
function of $(am_{PS})^2$. On both lattices we find a linear behavior except for the smallest 
quark mass on the finer lattice where the computed mass lies exceptionally high. We therefore 
exclude this point in our chiral extrapolation [including the point changes the fit
results to $A=1.47(8)$, $B=0.44(12)$ with $\chi^2/d.o.f.=3.42$].

Next, we discuss the results for the excited vector meson channel which are shown in the lower 
plots of Fig. \ref{fig:excited}. We find that our results on the coarse lattice are somewhat 
problematic. We observe a very jumpy behavior of the meson masses as a function of $(am_{PS})^2$. 
A reason for this might be that we had to choose different operator combinations for the 
different sea-quark masses. This also makes the chiral extrapolation very difficult. We try a 
linear fit as the simplest choice. This leads to a value of $\chi^2/d.o.f. \approx 4$ which 
shows that the fit is not reliable. Thus, the result should not be taken too seriously. On 
the finer lattice, we again find that the result for the smallest quark mass lies exceptionally 
high. Thus, we exclude also this point in our chiral extrapolation [including the point in our fit
changes the results to $A=1.58(7)$, $B=0.59(11)$ with $\chi^2/d.o.f.=1.87$]. Since the operators
we use for the vectors couple also to a state which has $J=3$ in the continuum, we cannot exclude 
the possibility that we are seeing a $\rho_3$ meson here. However, our results for operators with
a \textit{minimum} $J$ of 3 give an even higher mass (see below).

\subsection{High spin}
\subsubsection{Effective masses}

Next we present our results for high spin and exotic mesons, where we used single correlation 
functions to extract ground state masses. To select appropriate ranges where we can fit a single 
exponential function of the form $A_0 e^{-m_0 t}$ we use effective mass plots from the folded 
correlators.

\begin{figure*}
\begin{center}
\includegraphics[width=0.6\textwidth,clip]{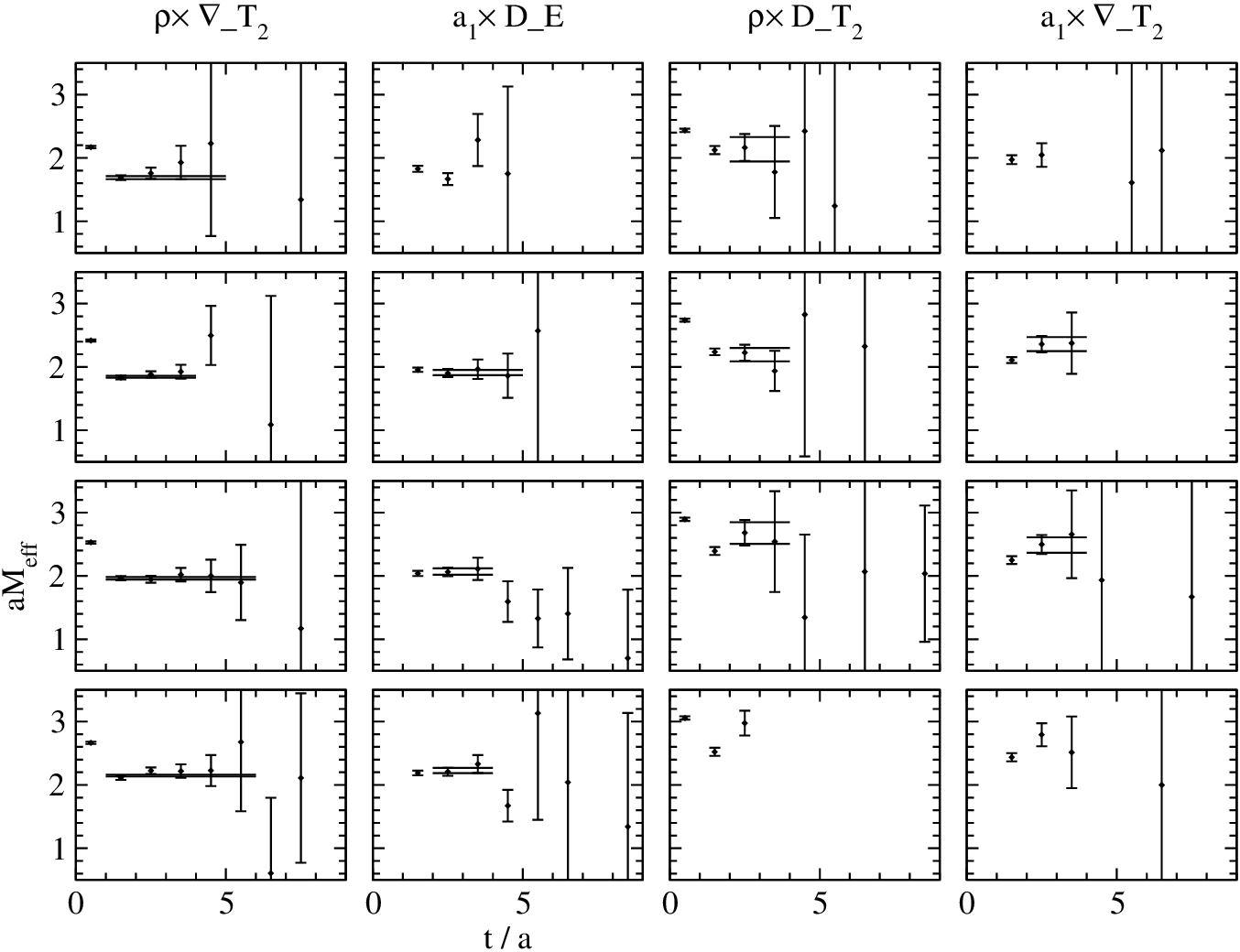}
\end{center}
\caption{Effective mass plots for the spin-2 meson interpolators $\rho\times\nabla\_T_2, a_1\times D\_E, 
\rho\times D\_T_2, a_1\times\nabla\_T_2$ from our coarse lattice (with $\kappa=0.1464,0.1445,0.1430,0.1409$ 
from top to bottom). Ground states are shown, along with the $M\pm\sigma_M$ results from correlated fits to 
the corresponding time intervals.}
\label{fig:eff-masses-12x24}
\end{figure*}

Figure \ref{fig:eff-masses-12x24} shows some selected plots from our coarse lattice, 
where we obtain signals for most of our operators coupling to spin $J=2$. 

For the $a_2$ meson which has quantum numbers $J^{PC}=2^{++}$ there are three operators 
available (see Ref.~\cite{Liao:2002rj}). Masses and fitting ranges for the interpolators 
$\rho\times\nabla\_T_2$ and $a_1\times D\_E$ are shown in Fig.~\ref{fig:eff-masses-12x24}. 
One can observe some short plateaus in time ranges $t=1-5$ and $2-5$ which are of a different 
quality for the various operators. Also the effective masses are not always consistent within 
the errors for the different operators, even though they should couple to the same state. 
Therefore, we fitted the two lowest lying plateaus belonging to the interpolators 
$\rho\times\nabla\_T_2$ and $a_1\times D\_E$.

Liao and Manke \cite{Liao:2002rj} also provide three operators that couple to $J^{PC}=2^{--}$. 
Their signal is weaker, the errors are somewhat bigger and it is often tricky to find 
appropriate plateaus. One may, for example, look at the plots for the interpolator 
$\rho\times D\_T_2$ and $a_1\times\nabla\_T_2$ in Fig.~\ref{fig:eff-masses-12x24}. However, we 
tried to fit the $\rho\times D\_T_2$ in a range $t=2-4$.

Our results for the $\pi_2$ meson which has quantum numbers $J^{PC}=2^{-+}$ are quite poor and 
noisy on the coarse lattice, so that we could not detect considerable plateaus.

Our correlated fit results for the interpolators which provide feasible signals on this coarse lattice are 
listed in Table \ref{tab:12x24-masses}.

\begin{figure*}
\begin{center}
\includegraphics[width=0.6\textwidth,clip]{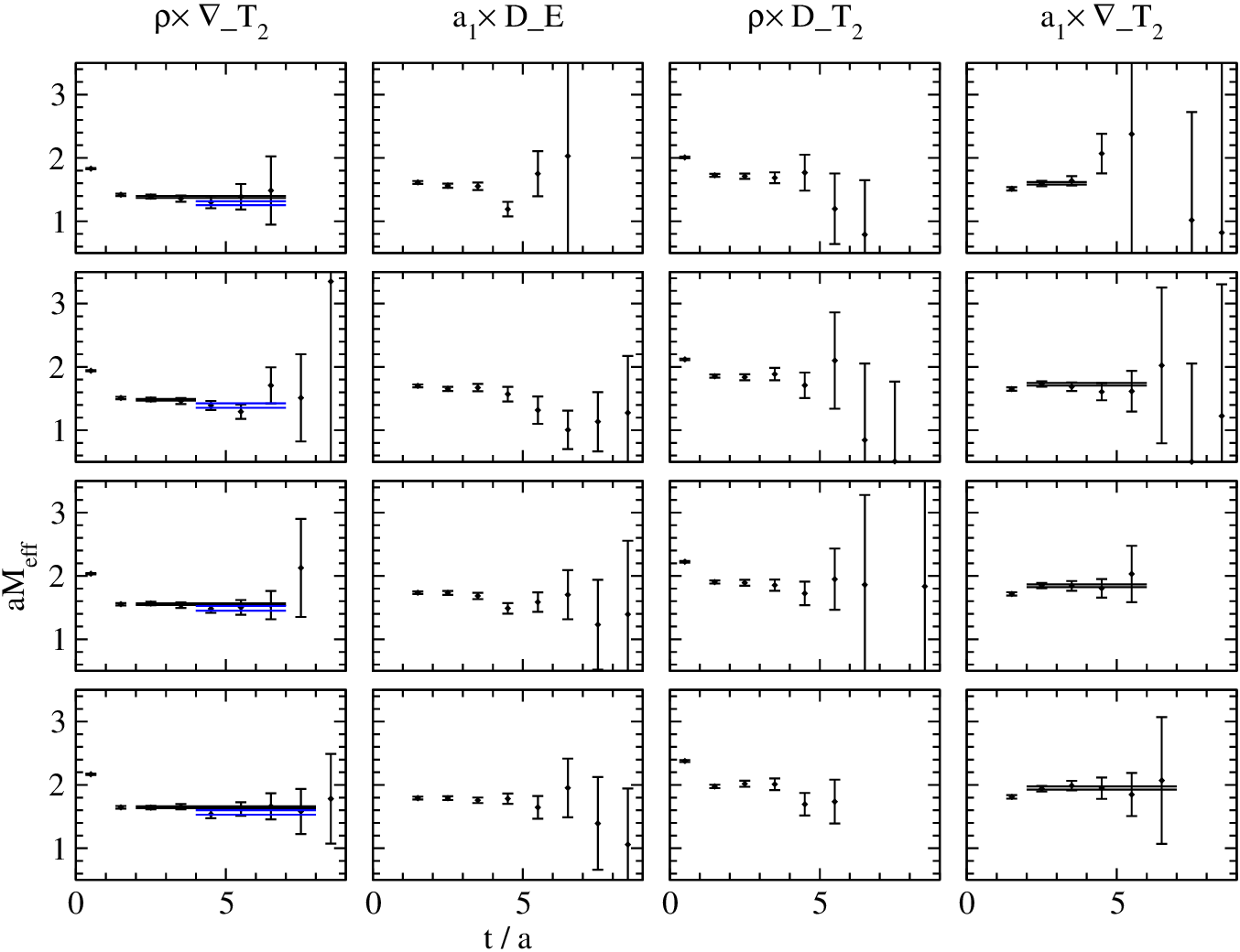}
\end{center}
\caption{Effective mass plots for the meson interpolators $\rho\times\nabla\_T_2, \rho\times D\_T_2, 
\rho\times D\_A_2, b_1\times\nabla\_T_1$ from our fine lattice (with $\kappa=0.1410,0.1400,0.1390,0.1375$ 
from top to bottom). Ground states are shown, along with the $M\pm\sigma_M$ results from correlated fits 
to the corresponding time intervals. For the $\rho\times\nabla\_T_2$ channel, we show results from two fitting 
ranges.}
\label{fig:eff-masses-16x32}
\end{figure*}

In Fig.~\ref{fig:eff-masses-16x32}, one can see example plots for interpolators with quantum numbers 
$J^{PC}=2^{++}$ and $2^{--}$ from our $16^3\times32$ lattice. 

Let us now have a closer look at these mesons on the finer lattice. For the $a_2$ meson we again 
plot the effective masses for the operators $\rho\times\nabla\_T_2$ and $a_1\times D\_E$. The 
plateaus have become longer and clearer but the discrepancies between the masses of the different 
interpolators have increased. For short distances $t < 4$ the plateaus for the $a_1\times D\_E$ 
and the $a_1\times D\_T_2$ are lying much higher than for the $\rho\times\nabla\_T_2$. Only for 
times $t > 5$ the masses become lower and agree within the errors, but simultaneously, the signal 
becomes very noisy.
Thus, we think that the $\rho\times\nabla\_T_2$ has more overlap with the ground state than the other two
operators, which might be contaminated strongly by excited states which then become suppressed only for large times.
Nevertheless, even for the $\rho\times\nabla\_T_2$ it is not unambiguous where to start fitting. This is why we 
present results from different fit ranges for this interpolator.

The situation for the $\rho_2$ state is similar to that of the $a_2$. In comparison to our coarse lattice one 
can observe reduced error bars, along with longer and clearer plateaus which are mostly consistant within the errors. Only 
the mass of the $\rho\times D\_T_2$ is somewhat higher. Moreover, there seems to be a step in the effective 
mass from the time slice $4$ to $5$ for this interpolator.
For this reason we fitted only the interpolators $a_1\times\nabla\_T_2$ and $a_1\times\nabla\_E$ for times $t > 2$.

We also observe very weak signals and huge errors for the $\pi_2$ meson on the $16^3\times32$ lattice. However, 
we tried to fit this state in time ranges $t=2-4$.

Even for some high spin mesons with spin quantum number $J=3$ and for the exotic $\pi_1$ state appropriate 
plateaus have been detected. Although their masses are quite high and the plateaus are really short, we present 
results for the interpolators $a_1\times D\_A_2, \rho\times D\_A_2$ and $b_1\times\nabla\_T_1$ in time ranges 
$t=1-4$ and $2-4$.

All our correlated fit results for the interpolators which provide sufficiently stable signals on this finer lattice are collected in 
Table \ref{tab:16x32-masses}.

\subsubsection{Chiral extrapolation}

For high spin and exotic meson states there are no results from chiral 
perturbation theory. However, we find that our masses depend almost linearly on $(am_{\pi})^2$. Therefore, we perform 
fits of the form
\begin{align}
am_{HS} = A_{HS} + B_{HS}(am_{PS})^2
\end{align}
to the physical point. The values for the physical point $am_{\pi}$, 
and the lattice spacing $a$ are listed in Table \ref{tab:physpoint}. 
Our extrapolation results are then summarized in Table \ref{tab:extrapol}.

\begin{figure*}
\begin{center}
\includegraphics[width=.4\textwidth,clip]{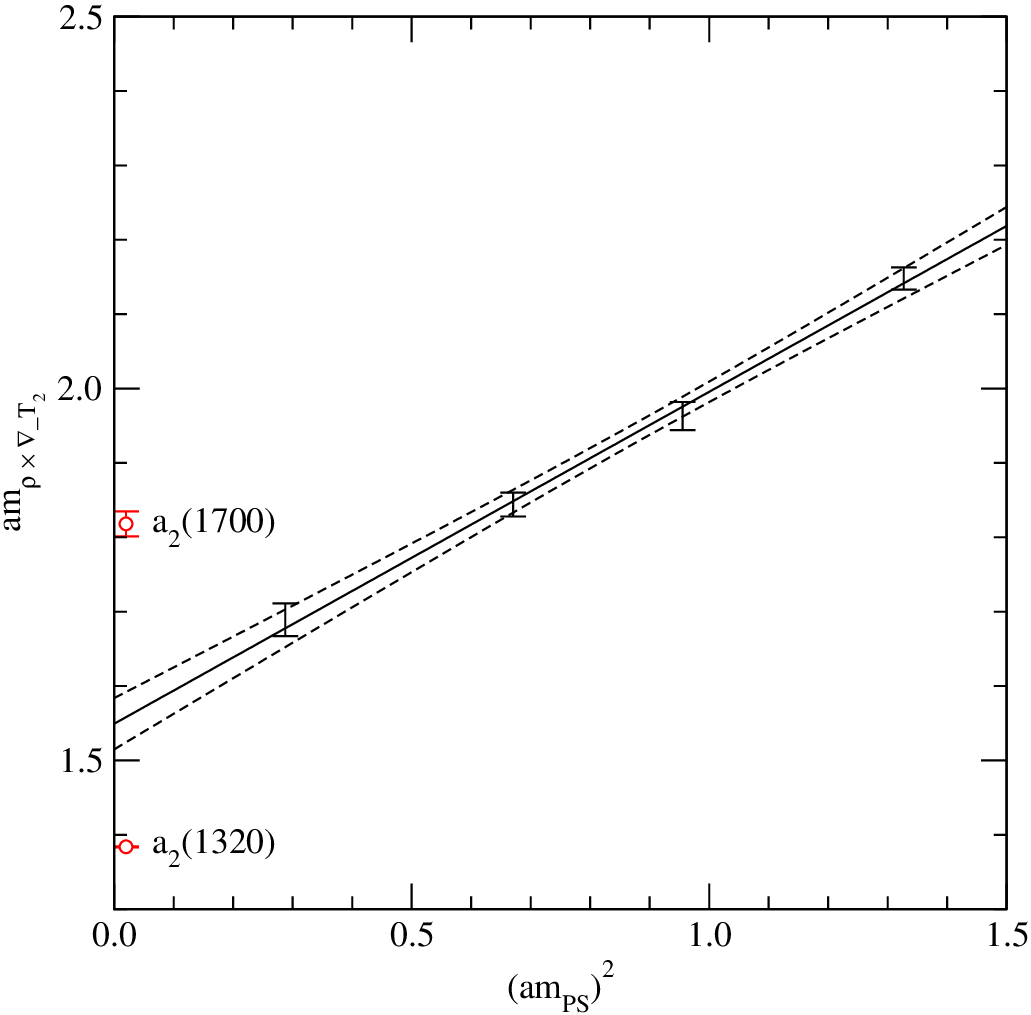}
\hspace*{5mm}
\includegraphics[width=.4\textwidth,clip]{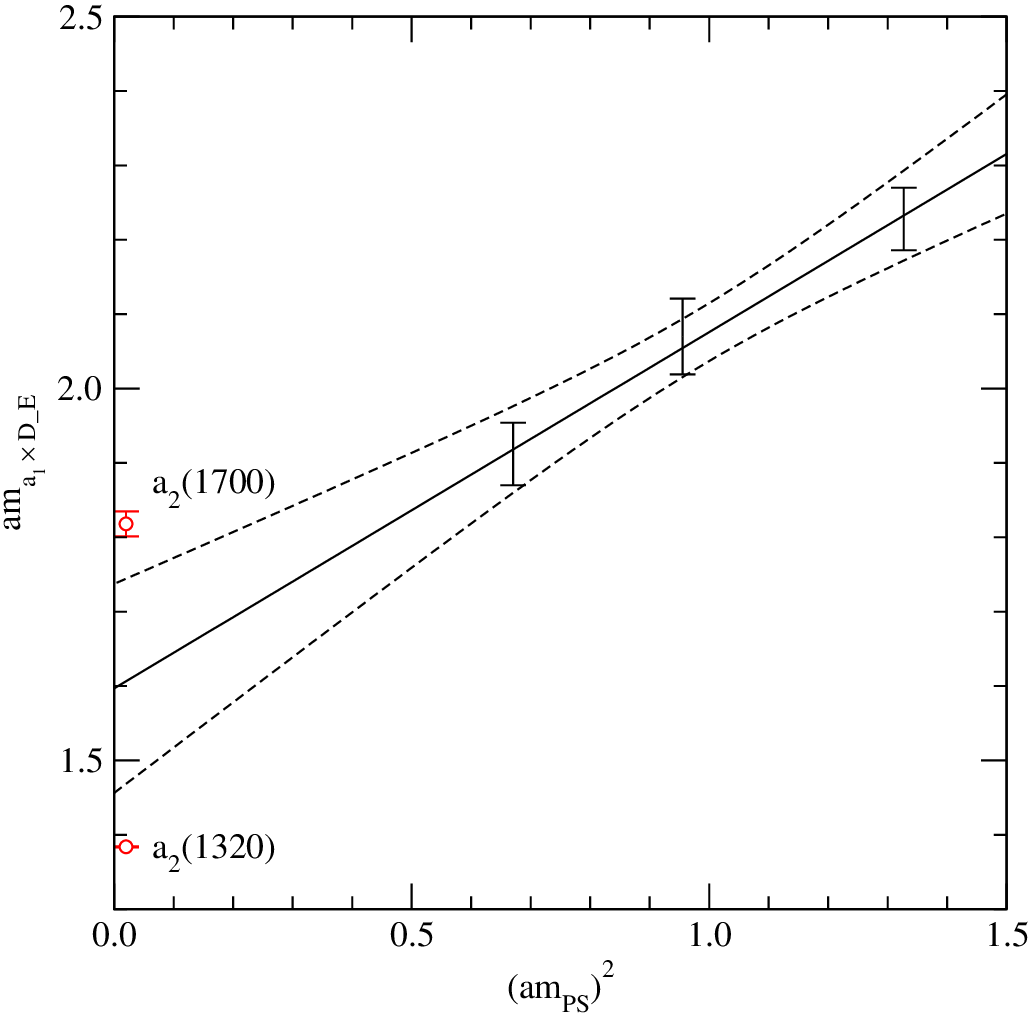}

\vspace*{5mm}

\includegraphics[width=.4\textwidth,clip]{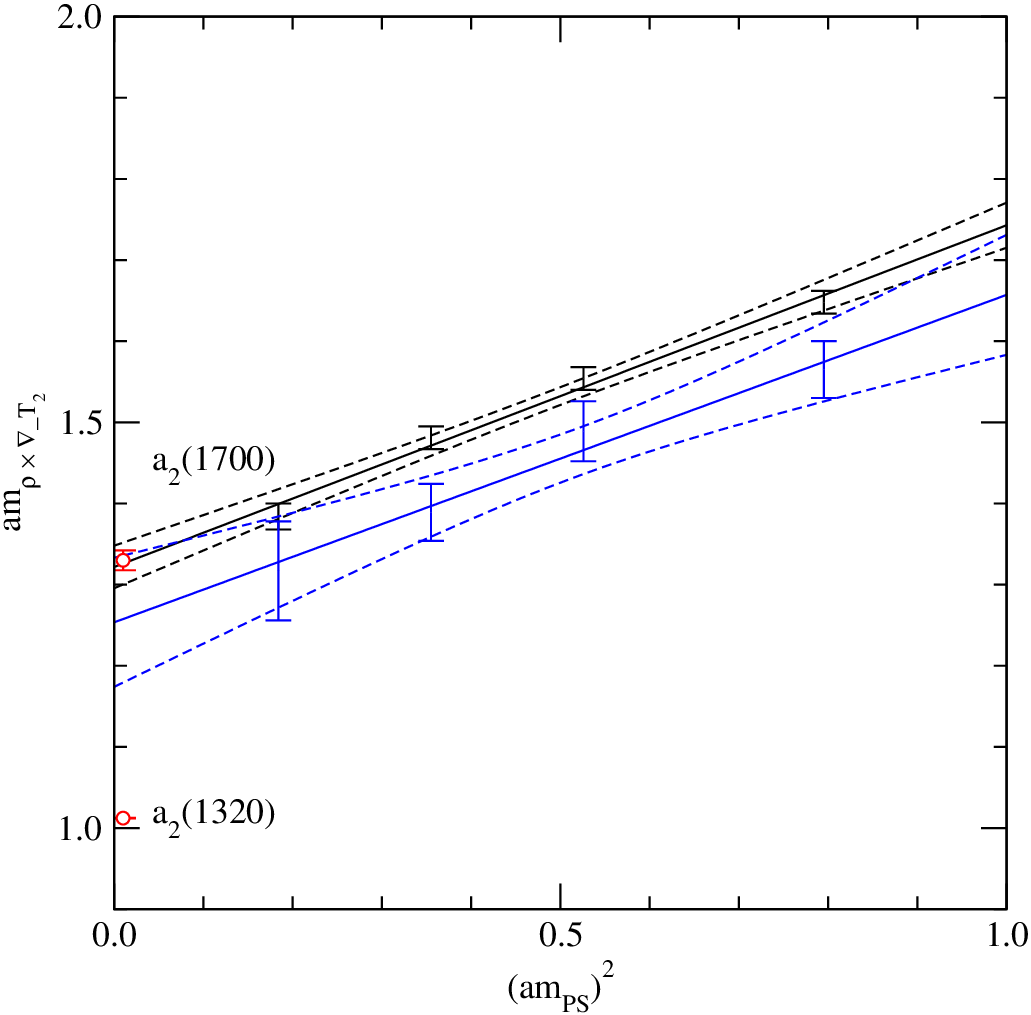}
\end{center}
\caption{The figure shows the ground states for our interpolators coupling to $J^{PC} = 2^{++}$ as a 
function of $(am_\pi)^2$. The upper plots are for the $12^3\times24$ lattice. They show the results for 
the interpolators $\rho\times\nabla\_T_2$ (left-hand side) and $a_1\times D\_E$ (right-hand side). The lower plot
is for the interpolator $\rho\times\nabla\_T_2$ on the $16^3\times32$ lattice and shows results for two
different fit ranges of the correlator (see Fig. \ref{fig:eff-masses-16x32}). The circles represent the experimental points. We also 
show the results of our chiral extrapolation (solid line) together with the one sigma error band 
(dashed lines).}
\label{fig:chiral-extrapol2++}
\end{figure*}

We first discuss the results for the $a_2$ meson channel which are shown 
in Fig. \ref{fig:chiral-extrapol2++}.
Here we notice a nice linear behavior of the operator 
$\rho\times\nabla\_T_2$ with small errors on the coarse lattice.
On the fine lattice however, we have two data sets belonging to two 
different ranges in the effective mass. We fit these
sets separately and average in the end, which introduces a possibly large systematic 
error.
For the $a_1\times D\_E$ we only have three data points on the coarse 
lattice available. Nevertheless, we try a linear fit.
This however, leads to higher errors and a small value of $\chi^2/d.o.f. = 
0.12$.

\begin{figure*}
\begin{center}
\includegraphics[width=.4\textwidth,clip]{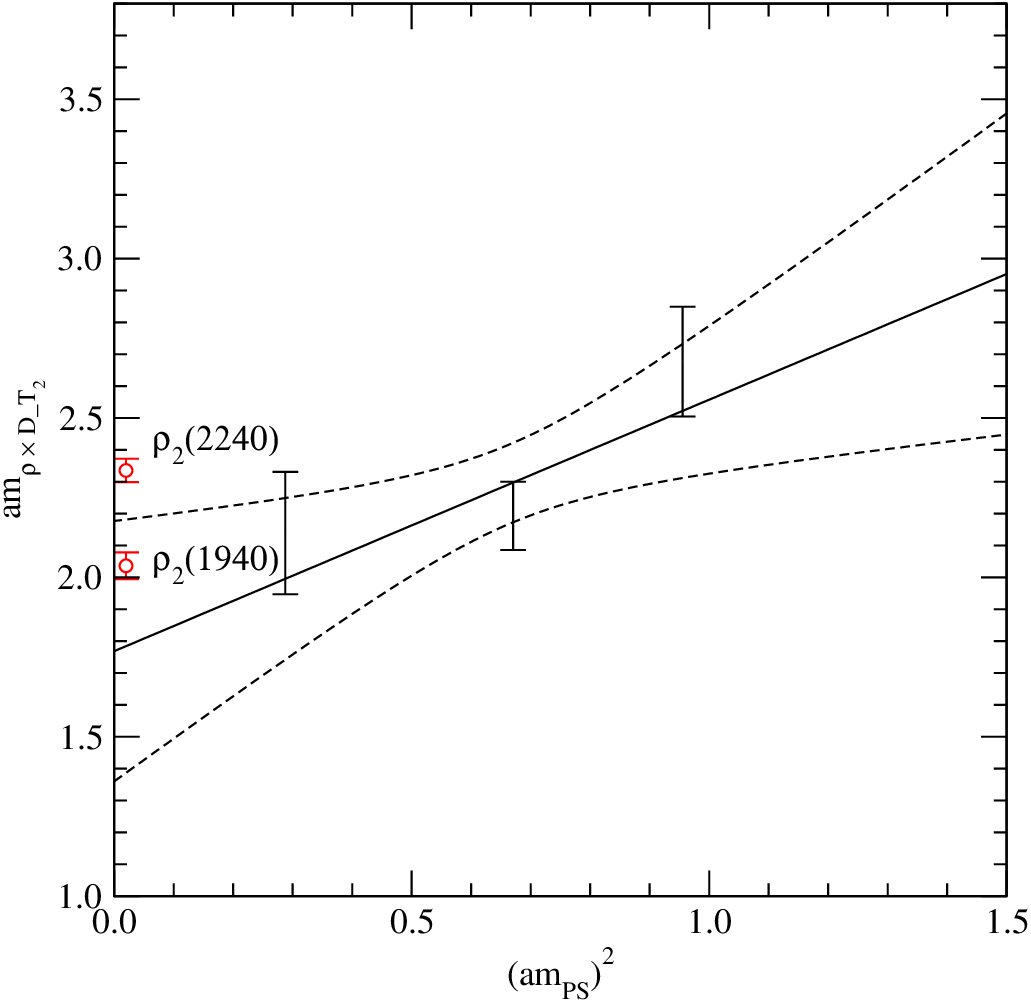}

\vspace*{5mm}

\includegraphics[width=.4\textwidth,clip]{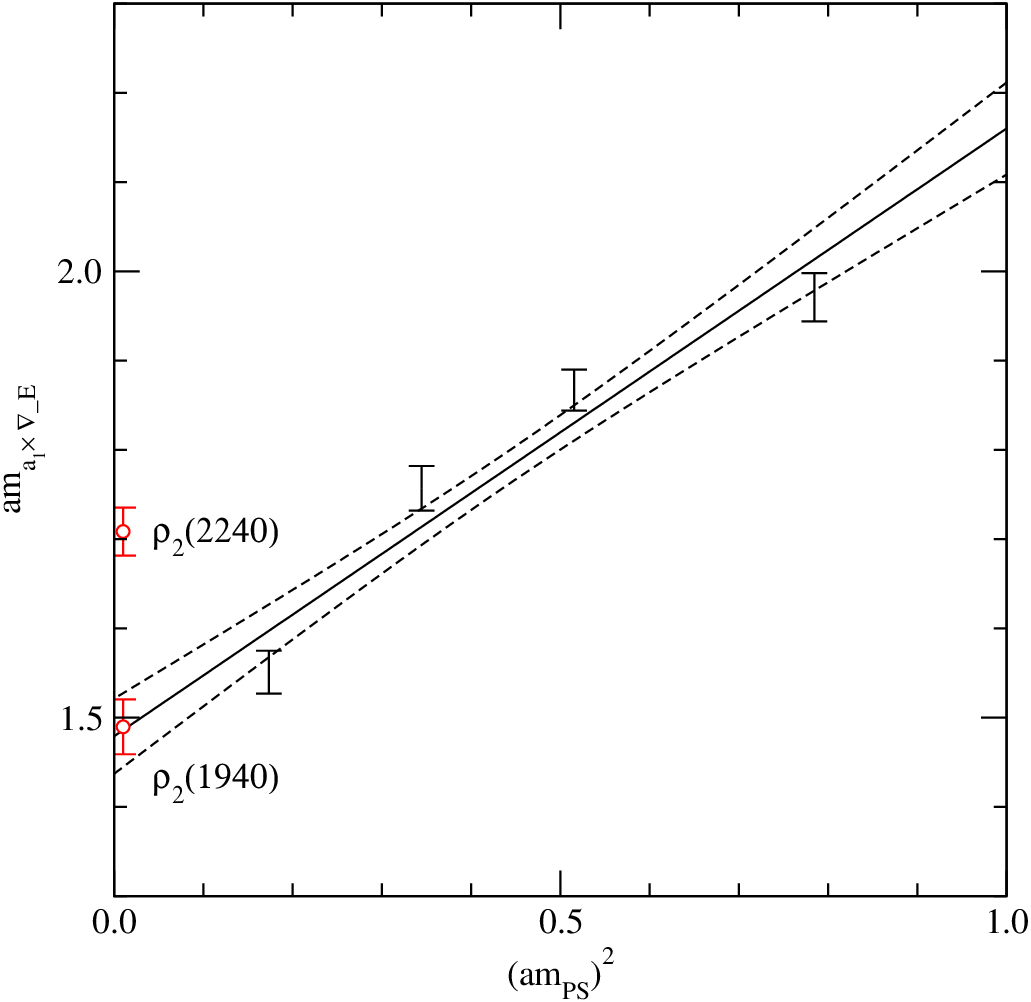}
\hspace*{5mm}
\includegraphics[width=.4\textwidth,clip]{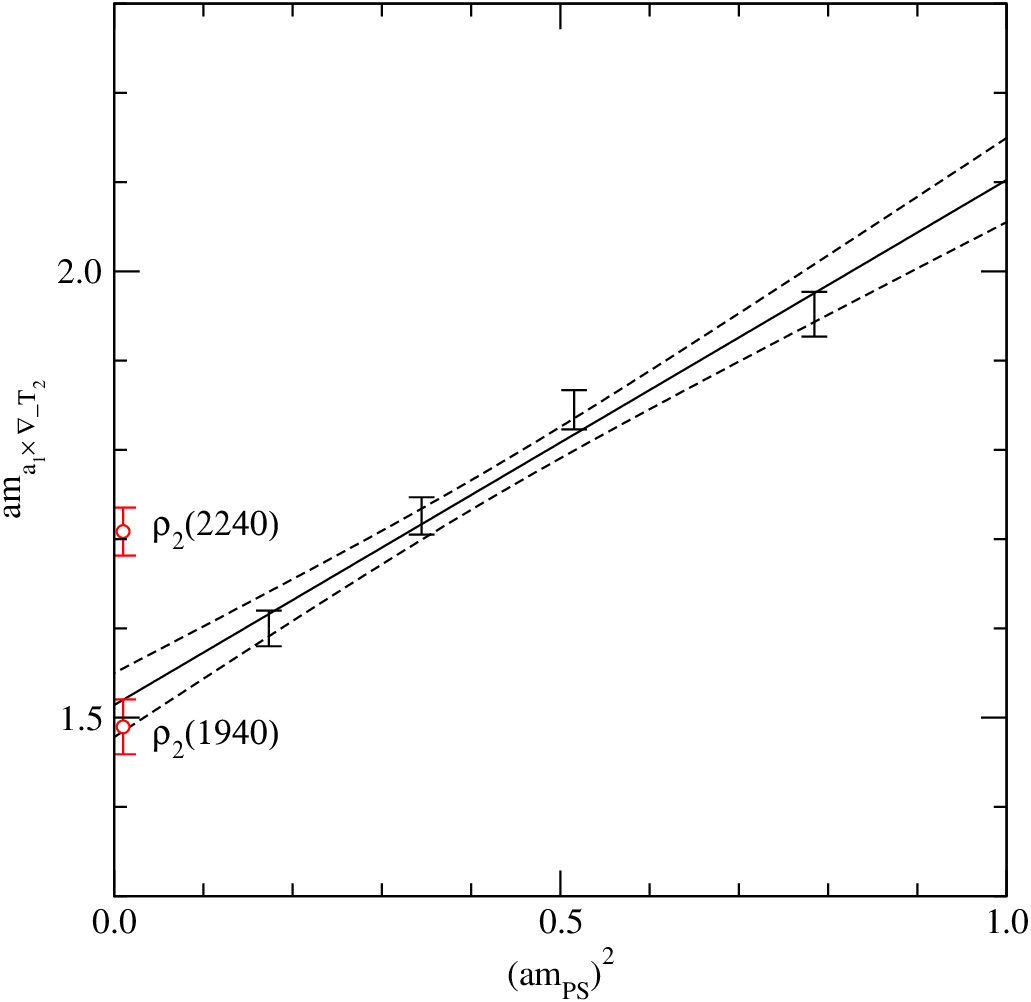}
\end{center}
\caption{The figure shows the ground states for our interpolators coupling to $J^{PC} = 2^{--}$ as a 
function of $(am_\pi)^2$. The upper plot is for the $12^3\times24$ lattice. It shows the results for 
the interpolator $\rho\times D\_T_2$. The lower plots are for the interpolators 
$a_1\times\nabla\_E$(left-hand side) and $a_1\times\nabla\_T_2$(right-hand side) on the $16^3\times32$ lattice.
The circles represent the experimental points. We also 
show the results of our chiral extrapolation (solid line) together with the one sigma error band 
(dashed lines).}
\label{fig:chiral-extrapol-2--}
\end{figure*}

In Fig. \ref{fig:chiral-extrapol-2--} we present our extrapolation 
results for the $\rho_2$ meson.
For the $\rho\times D\_T_2$ we also only have three data points on the 
coarse lattice. Additionally they show a
quite jumpy behavior of the meson mass as a function of $(am_{PS})^2$. 
This might be caused by the noisy signal on
that lattice and the brevity of the plateaus. Therefore, one should not 
trust this result too much.
Here the situation becomes better on the fine lattice where the 
interpolator $a_1\times\nabla\_T_2$ shows a good
linear behavior which leads to a reliable fit with small errors. For the 
$a_1\times\nabla\_E$ the masses are more
jumpys and therefore, we obtain a quite high value of 
$\chi^2/d.o.f. \approx 6$, although it agrees with the result of the 
interpolator $a_1\times\nabla\_T_2$.

\begin{figure*}
\begin{center}
\includegraphics[width=.4\textwidth,clip]{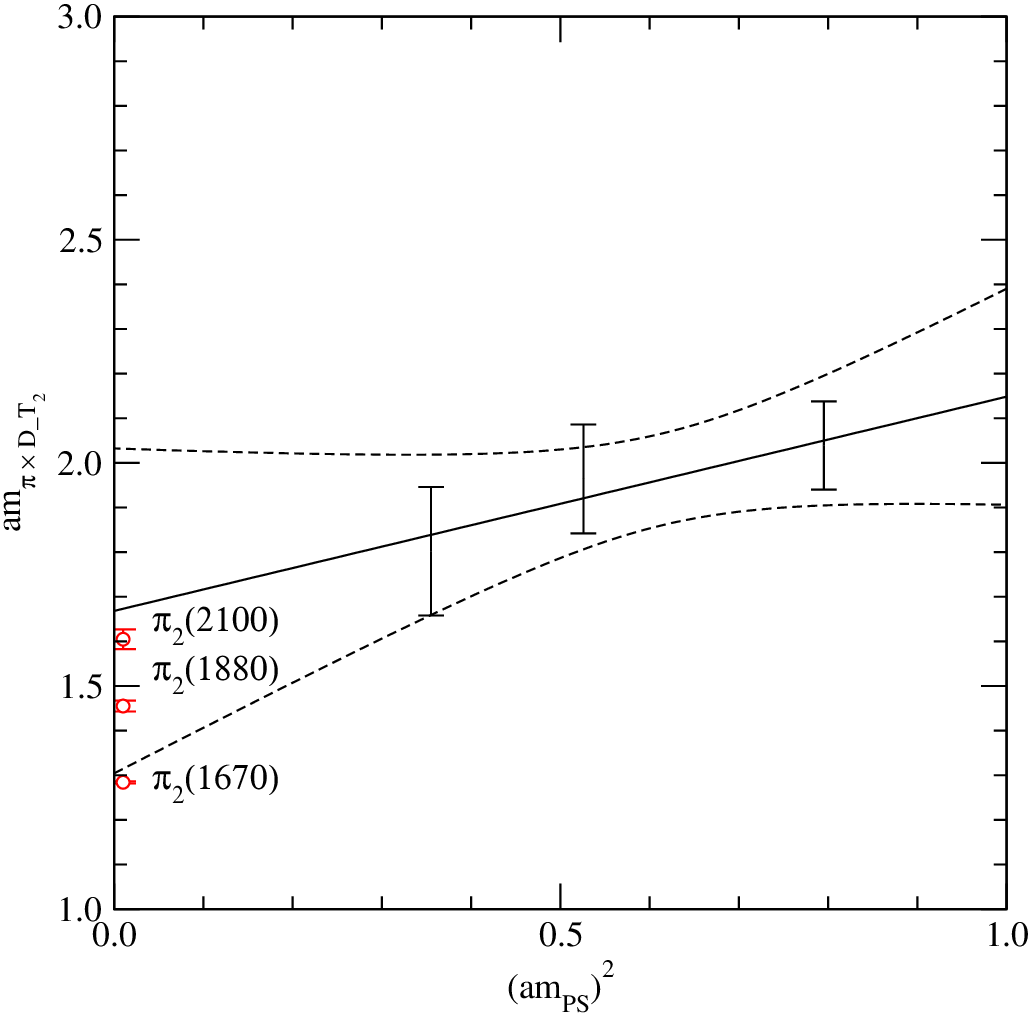}
\hspace*{5mm}
\includegraphics[width=.4\textwidth,clip]{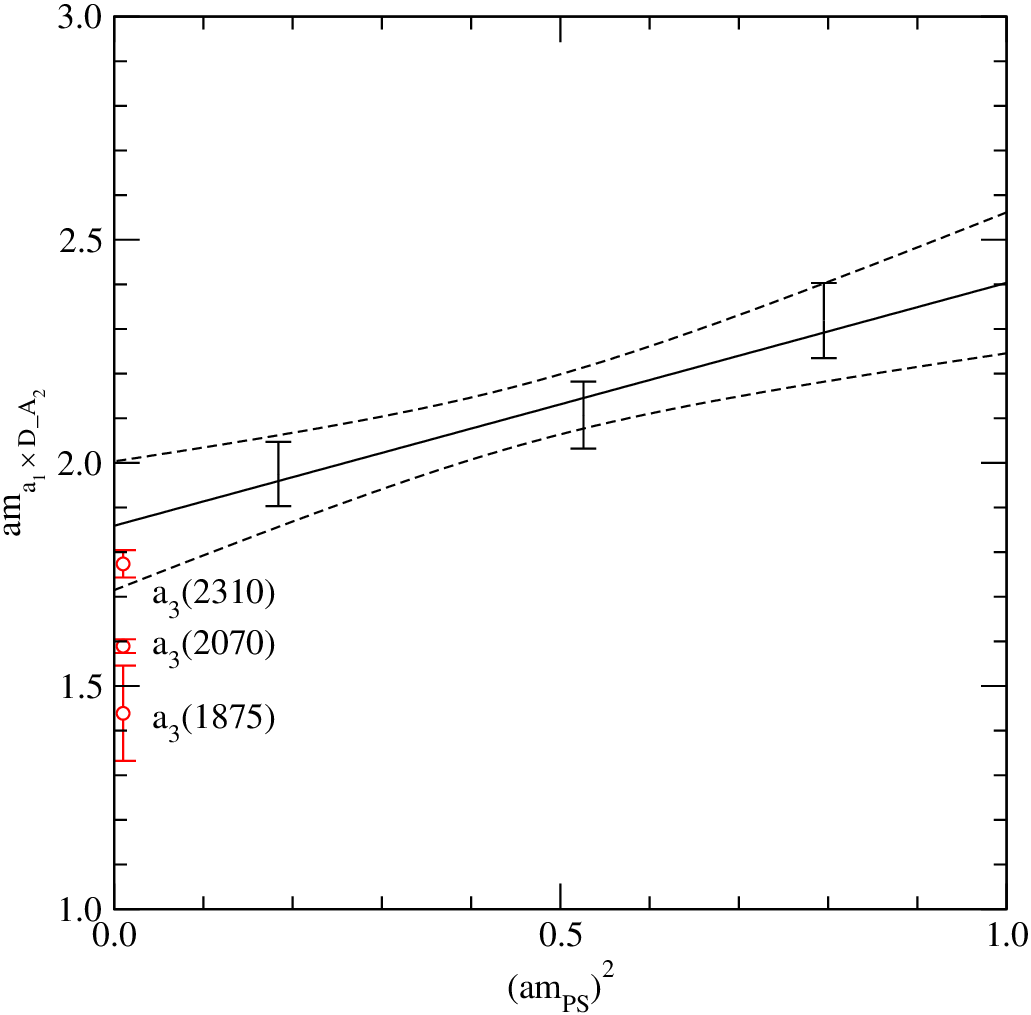}

\vspace*{5mm}

\includegraphics[width=.4\textwidth,clip]{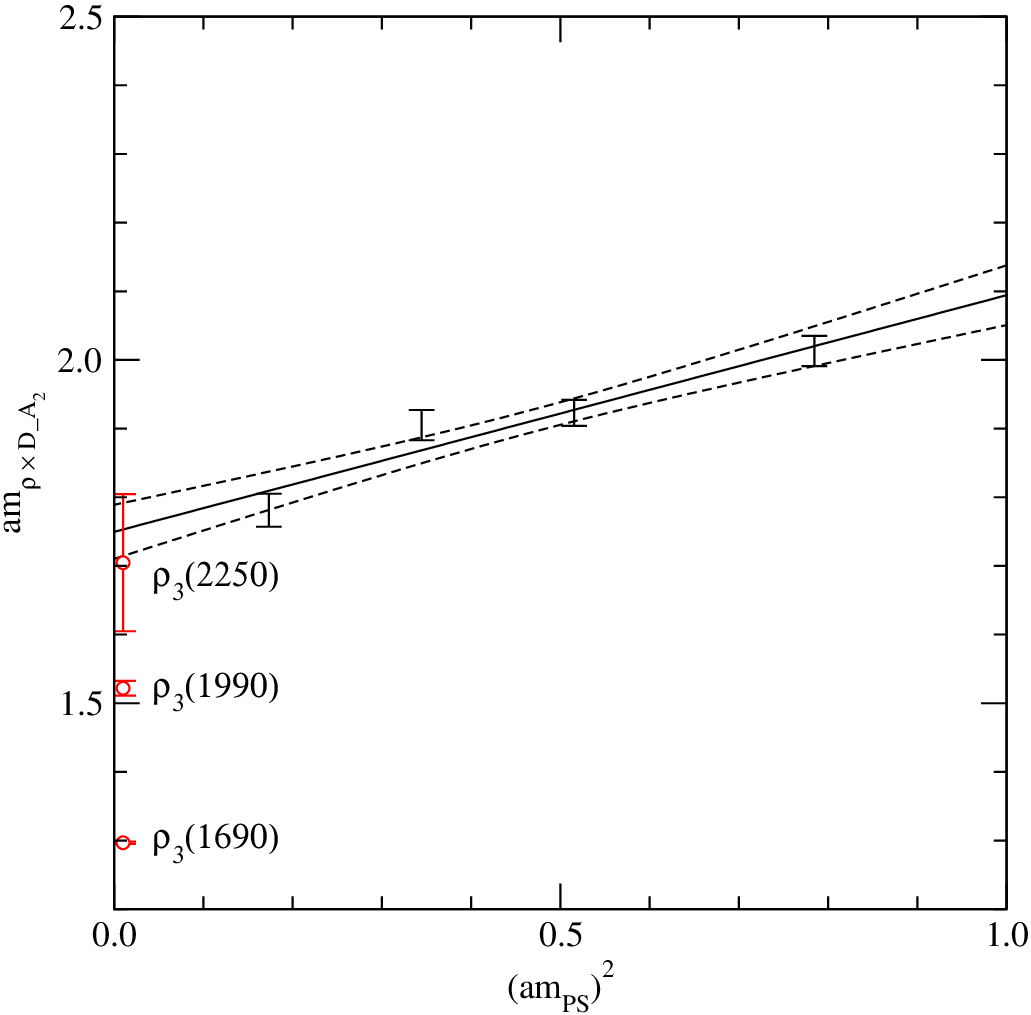}
\hspace*{5mm}
\includegraphics[width=.4\textwidth,clip]{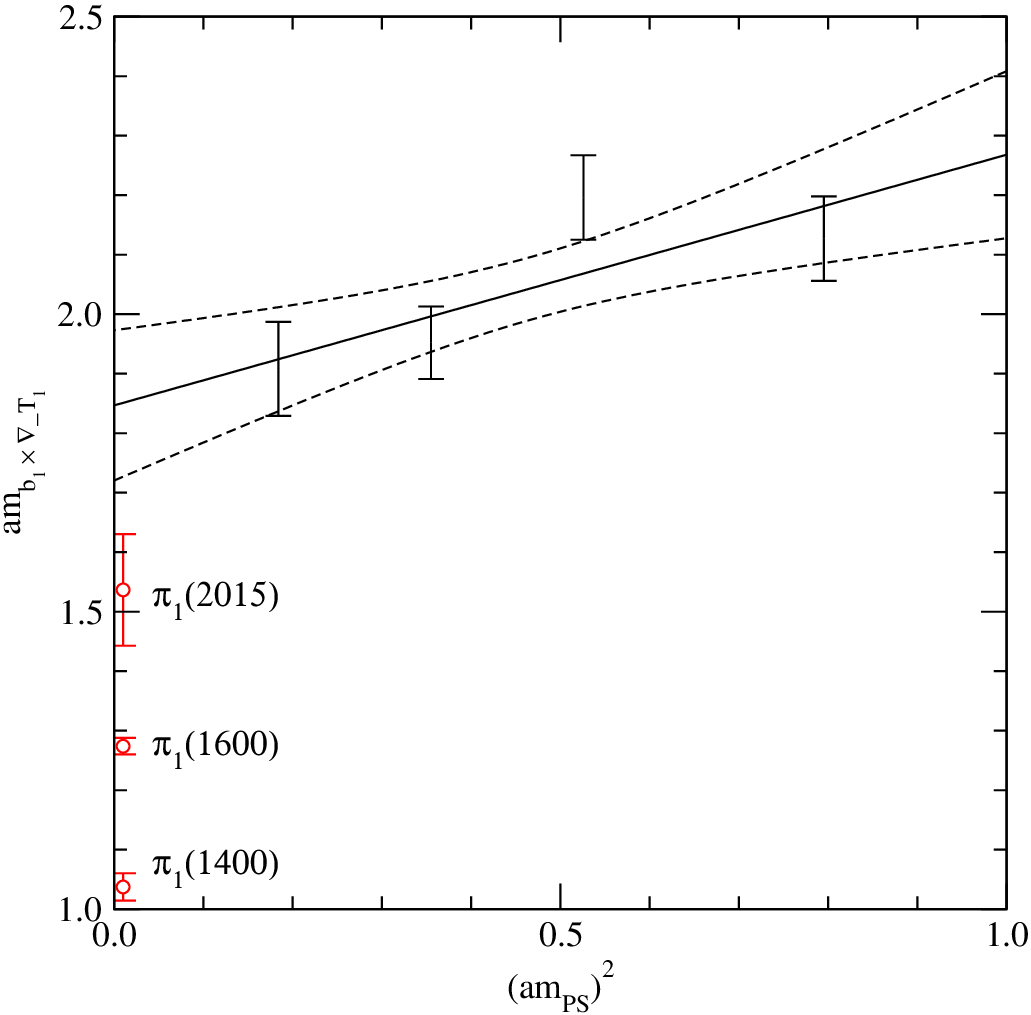}
\end{center}
\caption{The figure shows the ground states for interpolators which should couple to $J^{PC} = 2^{-+},3^{++},
3^{--},1^{-+}$ as a function of $(am_\pi)^2$. The quality of our data allows us to show only results 
for the $16^3\times32$ lattice for the interpolators $\pi\times D\_T_2, a_1\times D\_A_2, \rho\times D\_A_2$ 
and $b_1\times\nabla\_T_1$.
The circles represent the experimental points. We also 
show the results of our chiral extrapolation (solid line) together with the one sigma error band 
(dashed lines).}
\label{fig:chiral-extrapol-rest}
\end{figure*}

Finally we discuss the results for the $\pi_2, a_3,\rho_3$ and the exotic 
$\pi_1$  meson state which are shown
in Fig. \ref{fig:chiral-extrapol-rest} from top left to down right. For 
the operators $\pi\times D\_T\_2$ and
$a_1\times D\_A_2$ we again have only three data points available. 
Therefore, the value $\chi^2/d.o.f. < 0.5$.
For the operator $\rho\times D\_A_2$ we find a good linear behavior and 
obtain small errors.
The last plot shows the extrapolation of the interpolator 
$b_1\times\nabla\_T_1$ where we notice a small outlier
of the linear behavior at $(am_{PS})^2 \approx 0.53$. However, we also 
include this point into our fit because the
masses are afflicted with large errors.

\section{Discussion}

\subsection{Meson masses}

We compute the meson spectrum by evaluating the results of the chiral extrapolations at 
the physical point $am_\pi$ and then converting them into physical units by using our 
results for the lattice spacing $a$ (see Table \ref{tab:physpoint}). This means that for 
each meson channel we calculate
\begin{align}
M_{meson} & = \frac{[am_{meson}(am_\pi;A_{meson},B_{meson},C_{meson})]}{a},
\end{align}
where $A_{meson}$, $B_{meson}$, and $C_{meson}$ are the parameters that we have obtained 
from our chiral extrapolations and the $a$ in the denominator stands for the lattice 
spacing, which we have determined with the rho meson.

\begin{figure*}
\begin{center}
\resizebox{0.7\textwidth}{!}{
\includegraphics[clip]{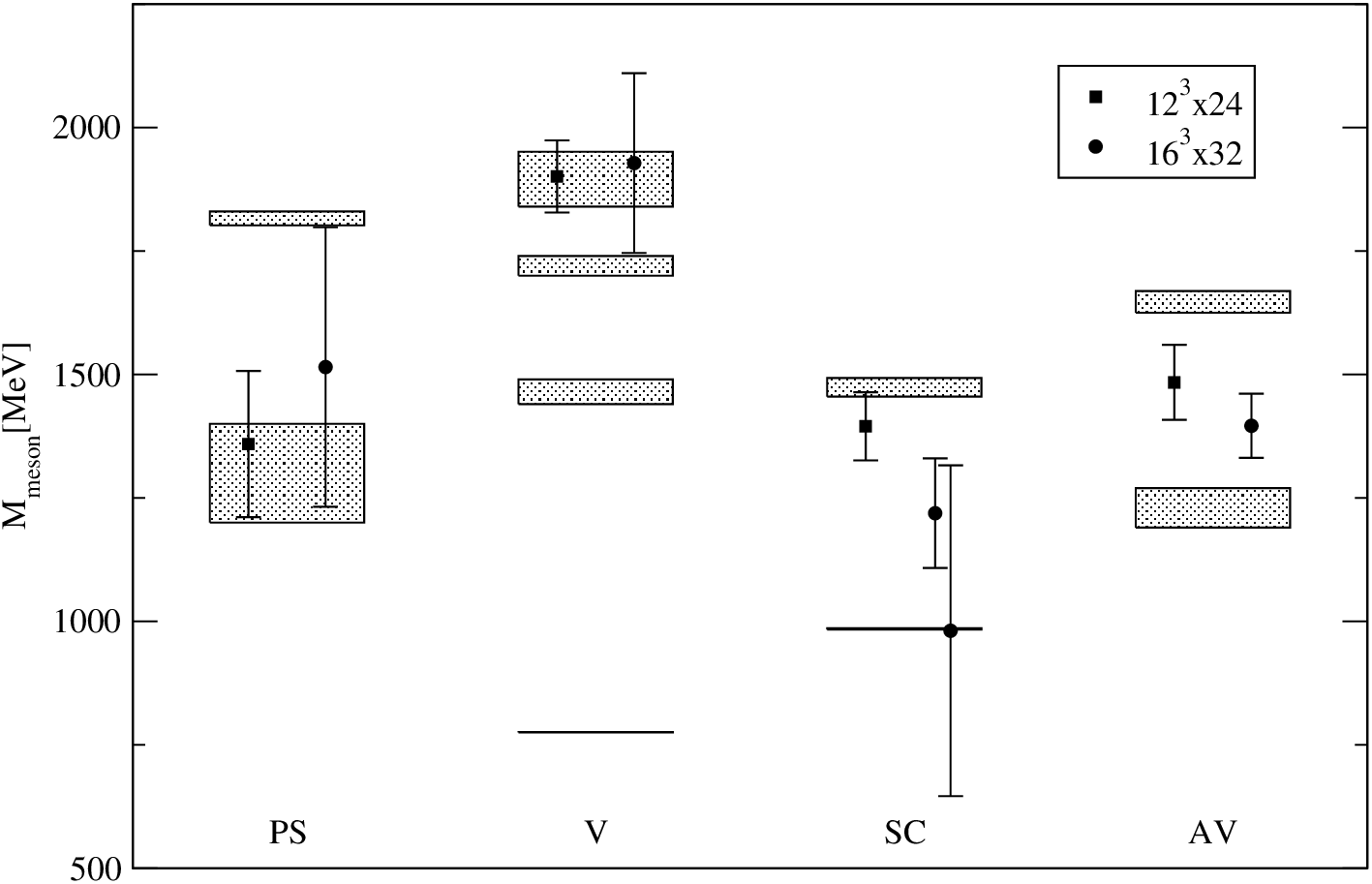}
}
\end{center}
\caption{Final results for the meson spectrum in the low spin sector. The 
boxes with the shaded areas represent the experimental values as classified 
by the Particle Data Group \cite{Amsler:2008zz}. For the scalar meson on the 
fine lattice we present results both for linear and quadratic extrapolation 
in $(am_{PS})^2$. The vector meson and pseudoscalar meson ground states are 
not shown, since the former is used to fix the lattice spacing $a$, and the 
latter becomes massless in the chiral limit.}
\label{fig:mesonfinals}
\end{figure*}

Our final results for the low spin meson spectrum are summarized in Fig. 
\ref{fig:mesonfinals}, where we plot our results for both lattices in comparison with 
the experimental values from \cite{Amsler:2008zz}. We do not show the vector meson ground 
state results since they have been used to determine the lattice spacing.

For the excited pseudoscalar meson our findings are in good agreement with the $\pi(1300)$ 
although the error for the finer lattice is quite large, thus making it also compatible 
with the $\pi(1800)$.

The results for the excited vector meson lie much too high. A reason for this might be the 
following: Our correlators are rather noisy, i.e., our effective mass plateaus are short, 
thus it might be that we start our fits too early. Another explanation is that our quark 
masses are too large and a more sophisticated extrapolation is needed. Unfortunately this 
is not possible since we have too few data points. We also want to mention here that we have 
found something similar in our previous quenched studies \cite{Burch:2006dg} on a coarse 
lattice. There, a finer lattice was needed to obtain better results.

For the scalar meson our results on the coarse lattice are compatible with the $a_0(1450)$. 
However, on the finer lattice we find smaller values. The linear extrapolated results lie 
between the $a_0(980)$ and the $a_0(1450)$. When a quadratic fit is used, the average value 
for the mass becomes smaller but the error is much larger. The first finding is similar to 
what we already have observed in previous quenched studies with approximate chiral fermions. 
First studies with dynamical CI-fermions \cite{Frigori:2007wa} however obtain a value which 
is consistent with the $a_0(980)$. This suggests that chiral sea quarks play a crucial role 
for scalar mesons.

For the axialvector meson our results are also higher than expected. They lie right between 
the $a_1(1260)$ and the $a_1(1640)$. This is similar to what we have seen in our previous 
quenched studies. Probably, here chiral sea quarks are needed to improve the situation, too.

\begin{figure*}
\begin{center}
\resizebox{0.7\textwidth}{!}{
\includegraphics[clip]{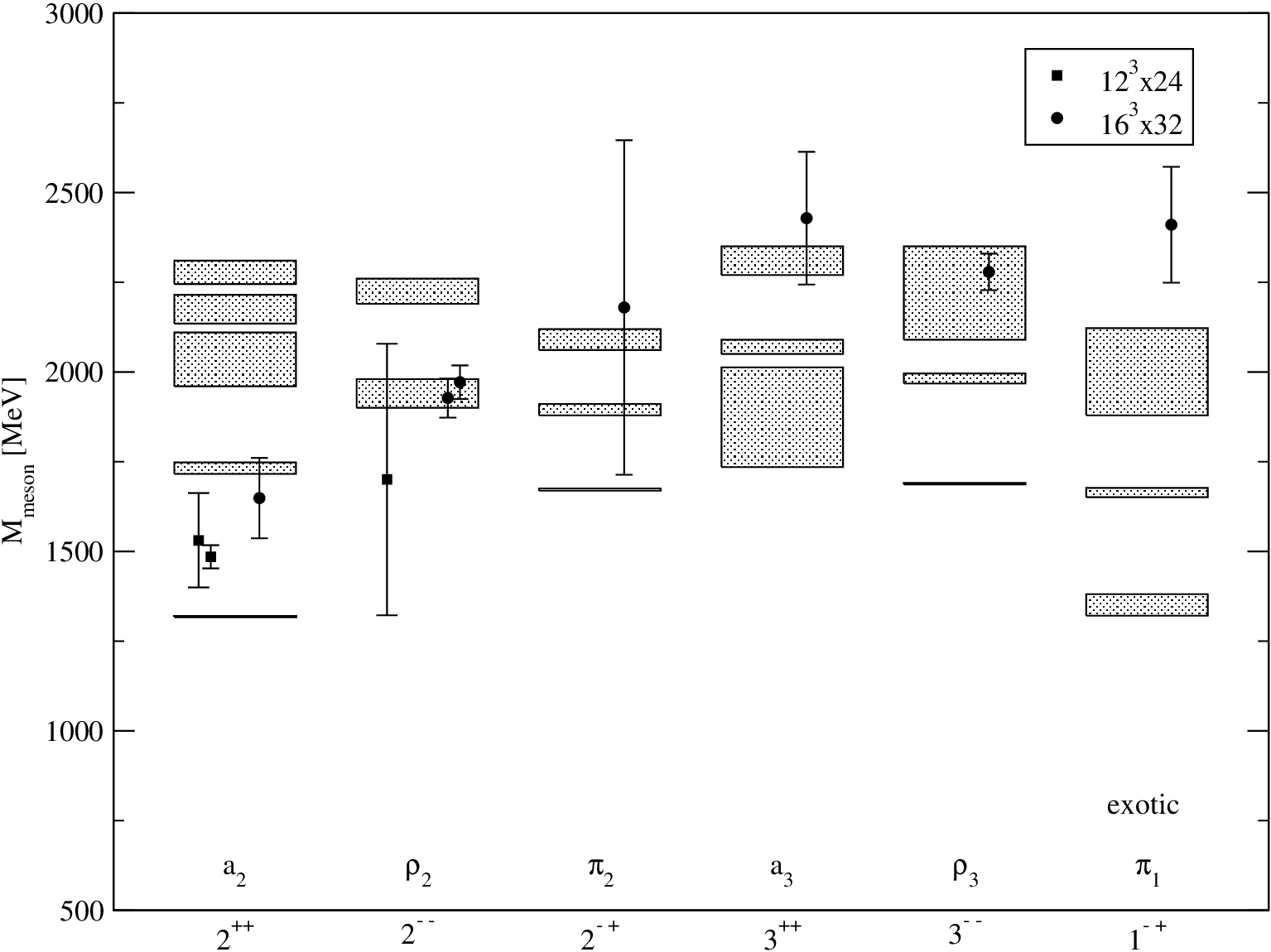}
}
\end{center}
\caption{Final results for the meson spectrum in the high spin and exotic 
sector. The boxes with the
shaded areas represent the experimental values as classified by the 
Particle Data Group \cite{Amsler:2008zz}. For the
$a_2$ meson on the coarse lattice and for the $\rho_2$ meson on the fine 
lattice we present results belonging
to different operators. For the $\pi_2$ meson, spin $J=3$ mesons and the 
exotic $\pi_1$ meson we only see
signals on the fine lattice.}
\label{fig:final results}
\end{figure*}

Our final results for the meson spectrum of high spin and exotic states 
are summarized in Fig. \ref{fig:final results},
where we again plot our results for both lattices in comparison to the 
experimental values from \cite{Amsler:2008zz}.

Our results for the $a_2$ meson lie between the $a_2(1320)$ and the 
$a_2(1700)$ which is higher than expected.
The effective masses for the discussed interpolators are quite short on 
the coarse lattice and also do not agree
within the errors for the various operators on both lattices. Therefore, 
it might be that the operators we used
have only poor overlap with the physical ground state and we start our 
fits too early. For this reason finer
lattices and a more advanced analysis, as it was done for the low spin 
mesons, would be needed to improve our
results. But also our usage of quite large quark masses, as mentioned 
above, might affect this shift.

For the $\rho_2$ meson we observe only weak signals and very short 
plateaus on the coarse lattice for one of
our operators coupling to that state. Thus we obtain quite large errors 
for our result on that lattice.
However, on the fine lattice we find very clear signals for that state and 
our results agree within the
errors and the physical ground state $\rho_2(1940)$.

For the $\pi_2$ meson we obtain only weak and noisy signals on the fine 
lattice. Hence, our result is afflicted with huge errors and lies too high.

We also found short effective mass plateaus for the high spin states 
$a_3$, $\rho_3$ and the exotic meson $\pi_1$.
The extrapolation, however, leads to masses much larger than those found for the 
experimental ground states. One possible
explanation for these findings might be finite volume effects, since 
these states should have more extended wave
functions. In this case, larger and finer lattices would be needed to obtain longer 
and clearer plateaus and to reduce discretization and finite volume effects.

\subsection{Possible systematics}

Since we work at pion masses above 500 MeV a number of hadronic decay channels which would
normally be open are suppressed. This introduces systematic shifts to our observed meson 
spectrum. Only by going to much smaller quark masses and taking into account explicit
mixing with multiparticle states  can we resolve these issues (see Refs. 
\cite{McNeile:2006bz,Cook:2006tz} for hybrid meson decays and Refs.
\cite{McNeile:2002fh,Aoki:2007rd,Gockeler:2008kc} for rho meson decay).

The lattices we use are about $2.5$fm in spatial extent. This may be exceptionally small for 
most of the excited mesons we study and may explain why many of our results come out too high.

We also use rather coarse lattices of $a=0.2$fm and $0.15$fm, making it difficult to 
unambiguously resolve the high masses of the excited states. 

It has been argued recently that there is a restoration of chiral symmetry in highly excited 
hadrons \cite{Glozman:2007ek}. Such considerations suggest that the use of (at least 
approximately) chiral fermions is important for lattice studies of excited states. Recent
efforts with dynamical chirally improved fermions \cite{Gattringer:2008vj} have appeared 
and work for the excited meson spectrum is in progress.

\begin{acknowledgments}
We thank Christof Gattringer and Christian B.~Lang for interesting discussions.
The calculations were performed on our local compute cluster and the QCDOC in 
Regensburg. We thank Tilo Wettig for supplying the machine and Stefan Solbrig 
for his help and technical support. We are also indepted to the CP-PACS 
Collaboration for sharing their configurations with us.
This work is supported in part by DFG (SFB-TR55) and GSI (RSCHAE). 
T.B. acknowledges previous support by the DOE(DE-FC02-01ER41183) and NSF(NSF-PHY-0555243).
\end{acknowledgments}

\begin{table*}
\begin{center}
\begin{tabular}{|c|c|c|c|c|}
\hline
$\kappa=\kappa_{sea}=\kappa_{val}$ & $am$ & $[t_{min},t_{max}]$ & $\chi^2/d.o.f.$ & Optimal operators  \\ \hline\hline
\multicolumn{5}{|c|}{Pseudoscalar ground state } \\ \hline\hline
$0.1409$ & 1.1520(23) & [5,10] & 0.16 & $P\gamma_5P$,$n\gamma_5n$,$\nabla_i\gamma_5\nabla_i$ \\ \hline
$0.1430$ & 0.9774(28) & [2,10] & 1.50 & $P\gamma_5P$,$n\gamma_5n$,$\nabla_i\gamma_5\nabla_i$ \\ \hline
$0.1445$ & 0.8201(29) & [2,9]  & 1.10 & $P\gamma_5P$,$n\gamma_5n$,$\nabla_i\gamma_5\nabla_i$ \\ \hline
$0.1464$ & 0.5363(60) & [2,9]  & 0.99 & $P\gamma_5P$,$n\gamma_5n$,$\nabla_i\gamma_5\nabla_i$ \\ \hline\hline
\multicolumn{5}{|c|}{Vector ground state} \\ \hline\hline
$0.1409$ & 1.4469(55) & [4,10] & 0.88 & $P\gamma_iP$,$L\gamma_iL$ \\ \hline
$0.1430$ & 1.3070(62) & [3,10] & 0.37 & $n\gamma_in$,$\nabla_i n$ \\ \hline
$0.1445$ & 1.1870(57) & [2,9]  & 0.35 & $P\gamma_iP$,$P\gamma_i\gamma_4P$,$n\gamma_in$,$L\gamma_iL$ \\ \hline
$0.1464$ & 0.973(15)  & [3,9]  & 0.41 & $P\gamma_iP$,$n\gamma_in$ \\ \hline\hline
\multicolumn{5}{|c|}{Scalar ground state} \\ \hline\hline
$0.1409$ & 2.188(28)  & [2,7]  & 0.17 & $\nabla_i\mathds{1}\nabla_i$ \\ \hline
$0.1430$ & 1.964(30)  & [2,5]  & 0.44 & $\nabla_i\mathds{1}\nabla_i$ \\ \hline
$0.1445$ & 1.824(29)  & [2,6]  & 0.84 & $\nabla_i\mathds{1}\nabla_i$ \\ \hline
$0.1464$ & 1.620(55)  & [2,5]  & 0.24 & $\nabla_i\mathds{1}\nabla_i$ \\ \hline\hline
\multicolumn{5}{|c|}{Axialvector ground state} \\ \hline\hline
$0.1409$ & 2.291(51)  & [3,6]  & 0.12 & $\nabla_i\gamma_k\gamma_5\nabla_i$ \\ \hline
$0.1430$ & 2.022(23)  & [2,6]  & 0.14 & $\nabla_i\gamma_k\gamma_5\nabla_i$ \\ \hline
$0.1445$ & 1.922(21)  & [2,6]  & 0.94 & $\nabla_i\gamma_k\gamma_5\nabla_i$ \\ \hline
$0.1464$ & 1.651(66)  & [3,6]  & 0.39 & $\nabla_i\gamma_k\gamma_5\nabla_i$ \\ \hline\hline
\multicolumn{5}{|c|}{Pseudoscalar 1st excited state} \\ \hline\hline
$0.1409$ & 2.276(40)  & [2,5]  & 0.24 & $P\gamma_5P$,$n\gamma_5n$,$\nabla_i\gamma_5\nabla_i$ \\ \hline
$0.1430$ & 2.003(90)  & [2,5]  & 0.16 & $P\gamma_5P$,$n\gamma_5n$,$\nabla_i\gamma_5\nabla_i$ \\ \hline
$0.1445$ & 1.868(62)  & [2,4]  & 0.01 & $P\gamma_5P$,$n\gamma_5n$,$\nabla_i\gamma_5\nabla_i$ \\ \hline
$0.1464$ & 1.56(14)   & [2,4]  & 0.71 & $P\gamma_5P$,$n\gamma_5n$,$\nabla_i\gamma_5\nabla_i$ \\ \hline\hline
\multicolumn{5}{|c|}{Vector 1st excited state} \\ \hline\hline
$0.1409$ & 2.436(50)  & [3,5]  & 0.01 & $P\gamma_iP$,$L\gamma_iL$ \\ \hline
$0.1430$ & 2.35(13)   & [2,4]  & 0.20 & $n\gamma_in$,$\nabla_i n$ \\ \hline
$0.1445$ & 2.082(48)  & [2,5]  & 0.53 & $P\gamma_iP$,$P\gamma_i\gamma_4P$,$n\gamma_in$,$L\gamma_iL$ \\ \hline
$0.1464$ & 2.128(42)  & [2,4]  & 0.10 & $P\gamma_iP$,$n\gamma_in$ \\ \hline
\end{tabular}
\end{center}
\caption{Results of the meson masses from the $12^3\times 24$ lattice. The interval $[t_{min},t_{max}]$ denotes the time range where we have fitted the eigenvalues. $\chi^2/d.o.f.$ represents the quality of our correlated fits. In the last column we show our final choice for the optimal operator combination for each meson channel.}
\label{tab:lowspinmasses12}
\end{table*}

\begin{table*}
\begin{center}
\begin{tabular}{|c|c|c|c|c|}
\hline
$\kappa=\kappa_{sea}=\kappa_{val}$ & $am$ & $[t_{min},t_{max}]$ & $\chi^2/d.o.f.$ & Optimal operators \\ \hline\hline
\multicolumn{5}{|c|}{Pseudoscalar ground state} \\ \hline\hline
$0.1375$ & 0.8917(24) & [4,13] & 0.58 & $P\gamma_5P$,$n\gamma_5n$,$\nabla_i\gamma_5\nabla_i$ \\ \hline
$0.1390$ & 0.7252(23) & [3,13] & 0.99 & $P\gamma_5P$,$n\gamma_5n$,$\nabla_i\gamma_5\nabla_i$ \\ \hline
$0.1400$ & 0.5958(22) & [5,11] & 0.88 & $P\gamma_5P$,$n\gamma_5n$,$\nabla_i\gamma_5\nabla_i$ \\ \hline
$0.1410$ & 0.4290(29) & [5,9]  & 0.81 & $P\gamma_5P$,$n\gamma_5n$ \\ \hline\hline
\multicolumn{5}{|c|}{Vector ground state} \\ \hline\hline
$0.1375$ & 1.1066(35) & [4,13] & 1.02 & $P\gamma_iP$,$n\gamma_in$ \\ \hline
$0.1390$ & 0.9648(48) & [4,12] & 0.80 & $P\gamma_iP$,$n\gamma_in$ \\ \hline
$0.1400$ & 0.8611(64) & [5,12] & 0.84 & $P\gamma_iP$,$n\gamma_in$ \\ \hline
$0.1410$ & 0.7332(82) & [5,13] & 0.40 & $P\gamma_iP$,$n\gamma_in$ \\ \hline\hline
\multicolumn{5}{|c|}{Scalar ground state} \\ \hline\hline
$0.1375$ & 1.583(41)  & [3,7]  & 0.43 & $L\mathds{1}n$ \\ \hline
$0.1390$ & 1.379(24)  & [2,8]  & 0.36 & $L\mathds{1}n$ \\ \hline
$0.1400$ & 1.263(34)  & [4,8]  & 0.17 & $\nabla_i\mathds{1}\nabla_i$ \\ \hline
$0.1410$ & 0.948(75)  & [4,7]  & 1.12 & $L\mathds{1}n$ \\ \hline\hline
\multicolumn{5}{|c|}{Axialvector ground state} \\ \hline\hline
$0.1375$ & 1.621(19)  & [2,7]  & 0.99 & $P\gamma_i\gamma_5P$,$n\gamma_i\gamma_5L$ \\ \hline
$0.1390$ & 1.334(74)  & [5,8]  & 0.12 & $P\gamma_i\gamma_5P$,$n\gamma_i\gamma_5L$ \\ \hline
$0.1400$ & 1.307(48)  & [4,8]  & 0.24 & $P\gamma_i\gamma_5P$,$n\gamma_i\gamma_5L$ \\ \hline
$0.1410$ & 1.199(28)  & [3,7]  & 0.39 & $P\gamma_i\gamma_5P$,$n\gamma_i\gamma_5L$ \\ \hline\hline
\multicolumn{5}{|c|}{Pseudoscalar 1st excited state} \\ \hline\hline
$0.1375$ & 1.838(30)  & [2,6]  & 0.34 & $P\gamma_5P$,$n\gamma_5n$,$\nabla_i\gamma_5\nabla_i$ \\ \hline
$0.1390$ & 1.605(62)  & [3,6]  & 0.37 & $P\gamma_5P$,$n\gamma_5n$,$\nabla_i\gamma_5\nabla_i$ \\ \hline
$0.1400$ & 1.46(12)   & [4,6]  & 0.01 & $P\gamma_5P$,$n\gamma_5n$,$\nabla_i\gamma_5\nabla_i$ \\ \hline
$0.1410$ & 1.660(74)  & [3,6]  & 0.03 & $P\gamma_5P$,$n\gamma_5n$ \\ \hline\hline
\multicolumn{5}{|c|}{Vector 1st excited state} \\ \hline\hline
$0.1375$ & 2.060(26)  & [3,5]  & 0.08 & $P\gamma_iP$,$n\gamma_in$ \\ \hline
$0.1390$ & 1.879(55)  & [4,7]  & 0.13 & $P\gamma_iP$,$n\gamma_in$ \\ \hline
$0.1400$ & 1.724(59)  & [4,7]  & 0.23 & $P\gamma_iP$,$n\gamma_in$ \\ \hline
$0.1410$ & 1.827(90)  & [4,6]  & 0.47 & $P\gamma_iP$,$n\gamma_in$ \\ \hline
\end{tabular}
\end{center}
\caption{The same as in Table \ref{tab:lowspinmasses12} but for the $16^3\times 32$ lattice.}
\label{tab:lowspinmasses16}
\end{table*}

\begin{table}
\begin{center}
\begin{tabular}{|c|c|c|c|c|}
\hline
\multicolumn{5}{|c|}{Pseudoscalar ground state} \\ \hline\hline
$L^3\times T$   & $\kappa_c^{-1}$ & B        & C         & $\chi^2/d.o.f.$ \\ \hline\hline
$12^3\times 24$ & 6.7678(28)      & 9.44(25) & -8.4(1.4) & 0.10 \\ \hline\hline
$16^3\times 32$ & 7.0366(22)      & 6.61(24) &  1.0(1.6) & 0.36 \\ \hline
\multicolumn{5}{c}{} \\ \hline
$L^3\times T$   & A         & B         & C          & $\chi^2/d.o.f.$ \\ \hline\hline
\multicolumn{5}{|c|}{Vector ground state} \\ \hline\hline
$12^3\times 24$ & 0.801(29) & 0.656(66) & -0.128(35) & 0.78 \\ \hline
$16^3\times 32$ & 0.586(18) & 0.857(74) & -0.255(67) & 0.37 \\ \hline\hline
\multicolumn{5}{|c|}{Scalar ground state} \\ \hline\hline
$12^3\times 24$ & 1.452(48) & 0.549(49) & ---         & 0.14 \\ \hline
$16^3\times 32$ & 0.927(57) & 0.85(11)  & ---         & 2.28 \\ \hline
$16^3\times 32$ & 0.73(14)  & 1.65(55)  & -0.75(51) & 2.41 \\ \hline\hline
\multicolumn{5}{|c|}{Axialvector ground state} \\ \hline\hline
$12^3\times 24$ & 1.546(54) & 0.528(63) & --- & 1.91 \\ \hline
$16^3\times 32$ & 1.064(24) & 0.696(53) & --- & 0.88 \\ \hline\hline
\multicolumn{5}{|c|}{Pseudoscalar 1st excited state} \\ \hline\hline
$12^3\times 24$ & 1.41(10)  & 0.652(32) & --- & 0.14  \\ \hline
$16^3\times 32$ & 1.15(15)  & 0.86(20)  & --- & 0.001 \\ \hline\hline
\multicolumn{5}{|c|}{Vector 1st excited state} \\ \hline\hline
$12^3\times 24$ & 1.988(52) & 0.309(63) & ---         & 3.94  \\ \hline
$16^3\times 32$ & 1.473(94) & 0.74(13)  & ---         & 0.13  \\ \hline
\end{tabular}
\end{center}
\caption{Numerical results of the chiral extrapolations of the different meson channels in Sec. \ref{lowspin}.}
\end{table}

\begin{table}
\begin{center}
\begin{tabular}{|c|c|c|c|c|}
\hline
$L^3\times T$   & $\kappa_{ud}^{-1}$ & $am_\pi$   & $am_\rho$ & $a$[fm]     \\ \hline\hline
$12^3\times 24$ & 6.7722(27)         & 0.1438(28) & 0.814(52) & 0.2071(132) \\ \hline\hline
$16^3\times 32$ & 7.0400(21)         & 0.1055(18) & 0.595(32) & 0.1515(82)  \\ \hline
\end{tabular}
\end{center}
\caption{Results for the redetermination of the physical point and the lattice spacing. In addition, 
the bare quark mass parameter $\kappa_{ud}^{-1}$ which corresponds to the mass of the up/down quark has been 
computed.}
\label{tab:physpoint}
\end{table}

\begin{table}
\centering
\begin{tabular}{|c|c|c|c|c|}
	\hline
	$\kappa=\kappa_{sea}=\kappa_{val}$ &  $am$  & $[t_{min},t_{max}]$ & $\chi^2 / d.o.f.$ & ops \\
	\hline\hline
	\multicolumn{5}{|c|}{$J^{PC}=2^{++}$} \\
	\hline \hline
	0.1409 & 2.148(15) & $[1,5]$ & 3.79 & $\rho\times\nabla\_T_2$ \\
	0.1430 & 1.963(19) & $[1,6]$ & 0.22 & $\rho\times\nabla\_T_2$ \\
	0.1445 & 1.844(16) & $[1,4]$ & 1.36 & $\rho\times\nabla\_T_2$ \\
	0.1464 & 1.689(22) & $[1,4]$ & 1.42 & $\rho\times\nabla\_T_2$ \\
	\hline\hline
	\multicolumn{5}{|c|}{$J^{PC}=2^{++}$} \\
	\hline \hline
	0.1409 & 2.228(41) & $[2,4]$ & 0.96 & $a_1\times D\_E$ \\
	0.1430 & 2.070(51) & $[2,4]$ & 0.08 & $a_1\times D\_E$ \\
	0.1445 & 1.912(42) & $[2,5]$ & 0.08 & $a_1\times D\_E$ \\
	0.1464 & ---       & ---     & ---  & $a_1\times D\_E$ \\
	\hline\hline
	\multicolumn{5}{|c|}{$J^{PC}=2^{--}$} \\
	\hline \hline
	0.1409 & ---        & ---     & ---   & $a_1\times\nabla\_T_2$ \\
	0.1430 & 2.487(121) & $[2,4]$ & 0.10  & $a_1\times\nabla\_T_2$ \\
	0.1445 & 2.360(111) & $[2,4]$ & 0.002 & $a_1\times\nabla\_T_2$ \\
	0.1464 & ---        & ---     & ---   & $a_1\times\nabla\_T_2$ \\
	\hline\hline
	\multicolumn{5}{|c|}{$J^{PC}=2^{--}$} \\
	\hline \hline
	0.1409 & ---       & ---     & ---  & $a_1\times\nabla\_E$ \\
	0.1430 & ---       & ---     & ---  & $a_1\times\nabla\_E$ \\
	0.1445 & 2.138(32) & $[1,4]$ & 0.21 & $a_1\times\nabla\_E$ \\
	0.1464 & 1.922(45) & $[1,4]$ & 0.46 & $a_1\times\nabla\_E$ \\
	\hline\hline
	\multicolumn{5}{|c|}{$J^{PC}=2^{--}$} \\
	\hline \hline
	0.1409 & ---        & ---     & ---  & $\rho\times D\_T_2$ \\
	0.1430 & 2.677(173) & $[2,4]$ & 0.03 & $\rho\times D\_T_2$ \\
	0.1445 & 2.193(107) & $[2,4]$ & 0.64 & $\rho\times D\_T_2$ \\
	0.1464 & 2.139(192) & $[2,4]$ & 0.20 & $\rho\times D\_T_2$ \\
	\hline
\end{tabular}
\caption{Meson masses from the $12^3\times24$ lattice.}
\label{tab:12x24-masses}
\end{table}

\begin{table}
\centering
\begin{tabular}{|c|c|c|c|c|}
	\hline
	$\kappa=\kappa_{sea}=\kappa_{val}$ &  $am$  & $[t_{min},t_{max}]$ & $\chi^2 / d.o.f.$ & ops \\
	\hline \hline
	\multicolumn{5}{|c|}{$J^{PC}=2^{++}$} \\
	\hline \hline
	0.1375 & 1.648(14) & $[2,8]$ & 1.64 & $\rho\times\nabla\_T_2$ \\
	0.1390 & 1.554(14) & $[2,7]$ & 0.93 & $\rho\times\nabla\_T_2$ \\
	0.1400 & 1.481(14) & $[2,4]$ & 0.93 & $\rho\times\nabla\_T_2$ \\
	0.1410 & 1.384(16) & $[2,7]$ & 0.54 & $\rho\times\nabla\_T_2$ \\
	\hline
	0.1375 & 1.565(35) & $[4,8]$ & 0.59 & $\rho\times\nabla\_T_2$ \\
	0.1390 & 1.489(37) & $[4,7]$ & 0.07 & $\rho\times\nabla\_T_2$ \\
	0.1400 & 1.389(35) & $[4,7]$ & 2.43 & $\rho\times\nabla\_T_2$ \\
	0.1410 & 1.317(61) & $[4,7]$ & 0.22 & $\rho\times\nabla\_T_2$ \\
	\hline \hline
	\multicolumn{5}{|c|}{$J^{PC}=2^{--}$} \\
	\hline \hline
	0.1375 & 1.952(25) & $[2,7]$ & 0.24 & $a_1\times\nabla\_T_2$ \\
	0.1390 & 1.845(22) & $[2,6]$ & 0.25 & $a_1\times\nabla\_T_2$ \\
	0.1400 & 1.726(21) & $[2,6]$ & 0.69 & $a_1\times\nabla\_T_2$ \\
	0.1410 & 1.600(20) & $[2,4]$ & 0.92 & $a_1\times\nabla\_T_2$ \\
	\hline \hline
	\multicolumn{5}{|c|}{$J^{PC}=2^{--}$} \\
	\hline \hline
	0.1375 & 1.971(27) & $[2,5]$ & 0.48 & $a_1\times\nabla\_E$ \\ 
	0.1390 & 1.866(23) & $[2,6]$ & 0.23 & $a_1\times\nabla\_E$ \\ 
	0.1400 & 1.757(25) & $[2,5]$ & 0.88 & $a_1\times\nabla\_E$ \\ 
	0.1410 & 1.551(24) & $[2,5]$ & 0.07 & $a_1\times\nabla\_E$ \\ 
	\hline \hline
	\multicolumn{5}{|c|}{$J^{PC}=2^{-+}$} \\
	\hline \hline
	0.1375 & 2.039(99)  & $[2,4]$ & 0.06 & $\pi\times D\_T_2$ \\
	0.1390 & 1.964(122) & $[2,4]$ & 0.25 & $\pi\times D\_T_2$ \\
	0.1400 & 1.802(144) & $[2,4]$ & 0.26 & $\pi\times D\_T_2$ \\
	0.1410 & ---        & ---     & ---  & $\pi\times D\_T_2$ \\
	\hline \hline
	\multicolumn{5}{|c|}{$J^{PC}=3^{++}$} \\
	\hline \hline
	0.1375 & 2.319(84) & $[2,4]$ & 1.11 & $a_1\times D\_A_2$ \\
	0.1390 & 2.107(75) & $[2,4]$ & 0.04 & $a_1\times D\_A_2$ \\
	0.1400 & ---       & ---     & ---  & $a_1\times D\_A_2$ \\
	0.1410 & 1.975(72) & $[2,4]$ & 0.02 & $a_1\times D\_A_2$ \\
	\hline \hline
	\multicolumn{5}{|c|}{$J^{PC}=3^{--}$} \\
	\hline \hline
	0.1375 & 2.013(22) & $[1,5]$ & 0.20 & $\rho\times D\_A_2$ \\
	0.1390 & 1.923(19) & $[1,4]$ & 0.19 & $\rho\times D\_A_2$ \\
	0.1400 & 1.905(22) & $[1,4]$ & 0.26 & $\rho\times D\_A_2$ \\
	0.1410 & 1.781(24) & $[1,4]$ & 0.15 & $\rho\times D\_A_2$ \\
	\hline\hline
	\multicolumn{5}{|c|}{$J^{PC}=1^{-+}$} \\
	\hline \hline
	0.1375 & 2.127(71) & $[2,4]$ & 0.0008 & $b_1\times\nabla\_T_1$ \\
	0.1390 & 2.196(71) & $[2,4]$ & 0.25 & $b_1\times\nabla\_T_1$ \\
	0.1400 & 1.952(61) & $[2,4]$ & 0.80 & $b_1\times\nabla\_T_1$ \\
	0.1410 & 1.908(79) & $[2,4]$ & 0.85 & $b_1\times\nabla\_T_1$ \\
	\hline
\end{tabular}
\caption{Meson masses from the $16^3\times32$ lattice.}
\label{tab:16x32-masses}
\end{table}

\begin{table}
\begin{center}
\begin{tabular}{|c|c|c|c|}
\hline
$L^3\times T$   & A         & B         & $\chi^2/d.o.f.$ \\ \hline\hline
\multicolumn{4}{|c|}{$2^{++} = \rho\times\nabla\_T_2$}    \\ \hline\hline
$12^3\times 24$ & 1.549(23) & 0.446(24) & 0.49            \\ \hline
$16^3\times 32$ & 1.322(17) & 0.421(33) & 1.18            \\ \hline
$16^3\times 32$ & 1.253(53) & 0.404(94) & 0.27            \\ \hline\hline
\multicolumn{4}{|c|}{$2^{++} = a_1\times D\_E$}           \\ \hline\hline
$12^3\times 24$ & 1.597(93) & 0.479(90) & 0.12            \\ \hline\hline
\multicolumn{4}{|c|}{$2^{--} = \rho\times D\_T_2$}        \\ \hline\hline
$12^3\times 24$ & 1.769(270)& 0.789(386)& 2.32            \\ \hline\hline
\multicolumn{4}{|c|}{$2^{--} = a_1\times\nabla\_T_2$}     \\ \hline\hline
$16^3\times32$ & 1.508(24)  & 0.588(50) & 1.67            \\ \hline\hline
\multicolumn{4}{|c|}{$2^{--} = a_1\times\nabla\_E$}       \\ \hline\hline
$16^3\times32$ & 1.472(28)  & 0.681(57) & 5.88            \\ \hline\hline
\multicolumn{4}{|c|}{$2^{-+} = \pi\times D\_T_2$}         \\ \hline\hline
$16^3\times32$ & 1.668(240) & 0.480(375)& 0.20            \\ \hline\hline
\multicolumn{4}{|c|}{$3^{++} = a_1\times D\_A_2$}         \\ \hline\hline
$16^3\times32$ & 1.859(95)  & 0.544(179)& 0.41            \\ \hline\hline
\multicolumn{4}{|c|}{$3^{--} = \rho\times D\_A_2$}        \\ \hline\hline
$16^3\times32$ & 1.746(26)  & 0.345(50) & 2.16            \\ \hline\hline
\multicolumn{4}{|c|}{$1^{-+} = b_1\times\nabla\_T_1$}     \\ \hline\hline
$16^3\times32$ & 1.847(83)  & 0.421(162)& 2.20            \\ \hline
\end{tabular}
\end{center}
\caption{Numerical results of the chiral extrapolations of the different high spin meson channels.}
\label{tab:extrapol}
\end{table}

\cleardoublepage

\end{document}